\documentclass[preprint,showpacs,preprintnumbers,amsmath,amssymb]{revtex4}
\usepackage[dvips]{epsfig}
\draft

\begin{document}
\overfullrule=0pt

\title{
\vskip-3cm
{\baselineskip14pt
\centerline{\normalsize\rm DESY 04--172 \hfill ISSN 0418--9833}
\centerline{\normalsize\rm MZ--TH/04--07 \hfill} 
\centerline{\normalsize\rm hep--ph/0410289 \hfill} 
\centerline{\normalsize\rm October 2004 \hfill}} 
\vskip1.5cm
\boldmath
Inclusive $D^{*\pm}$ Production in $p\bar{p}$ Collisions with Massive Charm
Quarks}

\author{B.A. Kniehl, G. Kramer, I. Schienbein} 

\affiliation{$II$. Institut f\"ur Theoretische Physik, Universit\"at Hamburg,
Luruper Chaussee 149, 22761 Hamburg, Germany}

\author{H. Spiesberger}

\affiliation{Institut f\"ur Physik, Johannes-Gutenberg-Universit\"at, 
Staudinger Weg 7, 55099 Mainz, Germany}

\thispagestyle{empty}

\begin{abstract}
We calculate the next-to-leading order cross section for the
inclusive production of $D^{*\pm}$ mesons in $p\bar{p}$ collisions as a
function of the transverse momentum and the rapidity in two
approaches using massive or massless charm quarks. For the inclusive
cross section, we derive the massless limit from the massive
theory. We find that this limit differs from the genuine massless
version with $\overline{\rm MS}$ factorization by finite corrections.
By adjusting
subtraction terms, we establish a massive theory with $\overline{\rm MS}$
subtraction which approaches the massless theory with increasing
transverse momentum. With these results and including the contributions
due to the charm and anti-charm content of the proton and anti-proton,
we calculate the inclusive $D^{*\pm}$ cross
section in $p \bar{p}$ collisions using realistic evolved non-perturbative
fragmentation functions and compare with recent data from the CDF
Collaboration at the Fermilab Tevatron at center-of-mass energy
$\sqrt{S} = 1.96$~TeV. We find
reasonable, though not perfect, agreement with the measured cross
sections.
\end{abstract}


\maketitle

\section{Introduction}

Recently, there has been quite some interest in the study of charm
production in proton--anti-proton collisions at high energies, both
experimentally and theoretically. The CDF Collaboration at the Fermilab
Tevatron presented results for prompt charm meson production cross
sections at center-of-mass energy $\sqrt{S}=1.96$ TeV. The differential cross
section $d\sigma/dp_T$ was measured as a function of transverse momentum
($p_T$) in the central rapidity ($y$) region $|y| \leq 1$
for inclusive production of $D^{0}$, $D^{+}$, $D^{*+}$ and $D_s^{+}$ mesons
and their charge conjugates \cite{CDF}.
For definiteness, we shall concentrate here on $D^{*\pm}$ mesons.
However, our results readily carry over to any other heavy-flavored hadrons.

On the theoretical side, various approaches for next-to-leading
order (NLO) calculations in perturbative QCD have been applied for
comparison with experimental data. In the so-called massless scheme 
\cite{CG,BKK}, also known as zero-mass variable-flavor-number (ZM-VFN) scheme,
which is the conventional
parton model approach implemented in the modified minimal-subtraction 
($\overline{\rm MS}$) scheme, the zero-mass parton approximation is applied
also to the charm quark, although its mass $m$ is certainly much larger
than the asymptotic scale parameter $\Lambda_{\rm QCD}$.
In this approach, the charm quark is also an incoming
parton originating from the proton or anti-proton, leading
to additional contributions, besides those from the gluon $g$ and the
$u$, $d$ and $s$ quarks.
The charm quark fragments into the $D^{*\pm}$ meson
similarly as the gluon and the light quarks with a fragmentation
function (FF) known from other processes.  The well-known factorization
theorem then provides a straightforward procedure for order-by-order
perturbative calculations.  Although this approach can be used as soon
as the factorization scales of the initial and final states are above
the starting scale of the parton distribution functions (PDFs) of the
(anti-)proton and of the FFs of the $D^{*\pm}$ meson, the predictions
are reliable only in the region of large transverse momenta $p_T \gg m$,
where terms of the order of $m^2/p_T^2$ can safely be neglected.

Another calculational scheme for heavy-flavor production which could be
applied to the process $p+\bar{p} \to D^{*\pm} + X$
\cite{DEN,BKNS,BMNSS,BS} is the so-called massive scheme, also called
fixed flavor-number (FFN) scheme, in which the number of active quark flavors
in the initial state is limited to $n_f=3$ and the charm quark
appears only in the final state. In this case, the charm quark is always
treated as a heavy particle and never as a parton. The actual mass
parameter $m$ is explicitly taken into account along with the variable $p_T$
as if they were of the same order, irrespective of their actual relative
magnitudes. In this scheme, the charm mass acts as a cutoff for the
initial- and final-state collinear singularities and sets the scale for
the perturbative calculations.  However, in NLO, terms proportional to
$\alpha_s(\mu_R)\ln(p_T^2/m^2)$, where $\mu_R$ is the renormalization scale,
arise from collinear emissions of a gluon
by the charm quark at large transverse momenta or from almost collinear
branchings of gluons into $c\bar{c}$ pairs. These terms are of order unity
for large values of $p_T$ and, with the choice $\mu_R \approx p_T$, they spoil
the convergence of the perturbation
series. The FFN approach with $n_f=3$ should thus be limited to a
rather narrow range of $p_T$ values, reaching up to a few times $m$.

There are also interpolating schemes, which smoothly interpolate between the
FFN scheme at low values of $p_T$ and the ZM-VFN scheme at large values of
$p_T$, with some freedom
concerning the detailed implementation.
The Aivazis-Collins-Olness-Tung (ACOT) \cite{acot} scheme, also known as
general-mass variable-flavor-number (GM-VFN) scheme, is one of them.
This scheme was applied to the hadroproduction of heavy flavors in
Ref.~\cite{ost} taking into account the FFN part at NLO and the leading
logarithms of the ZM-VFN part.

Another interpolating scheme which has been applied to inclusive $D^{*\pm}$
production in the Tevatron region is the so-called fixed-order
next-to-leading-logarithmic (FONLL) scheme, in which the traditional
cross section in the FFN scheme and a suitably modified cross section in
the ZM-VFN scheme with perturbative FFs are linearly
combined \cite{CGN,CN}. The combination is done in such a way, that the
ZM-VFN term is weighted with an ad-hoc coefficient function of the form
$p_T^2/(p_T^2+25 m^2)$ to enforce its suppression in the low-$p_T$
range.  In both finite-charm-mass approaches, the FFN and the FONLL, the
theoretically calculated FFN cross sections are convoluted with a
non-perturbative FF extracted from $e^{+} e^{-}$ data.  This assumes
universality of the FF which is not supported by a factorization theorem
as in the ZM-VFN approach.

As has been explained at many places in the literature, mainly in the
context of charm production in deep-inelastic $ep$ scattering (for a
recent review, see Ref.~\cite{TKS}), the correct approach for $p_T \gg m$
is to absorb the potentially large logarithms into the charm PDF of the
(anti-)proton and the FF of the $c\to D^{*+}$ transition.  Then,
large logarithms of the type $\ln(\mu_F^2/m^2)$, defined with the factorization
scale $\mu_F$, determine the evolution to higher scales and can be resummed
by virtue of the Dokshitzer-Gribov-Lipatov-Altarelli-Parisi (DGLAP) 
\cite{dglap} evolution equations. The unsubtracted terms of the form
$\ln(p_T^2/\mu_F^2)$ are of order unity for the appropriate
choice of $\mu_F$ of order $p_T$.  After factorizing the $\ln m^2$ terms,
the hard cross section is infrared safe, and $n_f=4$ is taken in
the formula for $\alpha_s$ and the DGLAP evolution equations. The remaining
dependence on $m$, {\it i.e.}\ the terms proportional to $m^2/p_T^2$, can be
kept in the hard cross section to achieve better accuracy in the
intermediate region $p_T \agt m$.  The factorization of mass-divergent terms
can be extended consistently to higher orders in
$\alpha_s$, as has been shown by Collins in the context of heavy-flavor
production in high-$Q^2$ $ep$ collisions \cite{Coll}.

It is well known that the subtraction of just the collinearly (mass)
singular terms does not define a unique factorization prescription.
Also finite terms must be specified.  In the conventional ZM-VFN
calculation, the mass $m$ is put to zero from the beginning and the
collinearly divergent terms are defined with the help of dimensional
regularization. This fixes the finite terms in a specific way, and their
form is inherent to the chosen regularization procedure.  If one starts
with $m \neq 0$ and performs the limit $m \to 0$ afterwards, the
finite terms can be different. These terms have to be removed by
subtraction together with the $\ln m^2$ terms in such a way that, in the
limit $p_T \to \infty$, the known ZM-VFN expressions
are recovered. This requirement is actually unavoidable, since almost all
existing PDFs and FFs, including those for heavy flavors, are defined
in this particular scheme (or sometimes in the deep-inelastic-scattering (DIS)
scheme, which can be derived from the $\overline{\rm MS}$ scheme).
It is clear that a subtraction scheme
defined in this way is a correct extension of the conventional
ZM-VFN scheme to include charm-quark mass effects in a consistent way.
In the following, we shall refer to it as the GM-VFN scheme, since it is
conceptionally similar to the framework of Ref.~\cite{ost}.
For a fully consistent analysis of heavy-flavor production in $p \bar{p}$
collisions, it will eventually be necessary to use dedicated PDFs and FFs
with heavy-quark mass effects included, determined by global fits
utilizing massive hard-scattering cross sections. Needless to say that
it is, therefore, important to work out massive hard-scattering
coefficients in one particular scheme for all relevant processes.
Actually, just recently PDFs of the proton with heavy-quark mass effects
included have been constructed by members of the CTEQ Collaboration
\cite{KLOT} in a scheme very similar to ours as outlined above.
If these were used
in a calculation of charm production in $p \bar{p}$ collisions, the
treatment of the corresponding hard-scattering cross sections would have
to be adjusted to these PDFs. However, we think that this would be
premature as long as similar constructions of the FF for $c \to
D^{*+}$ do not exist. 

In a recent work, two of us applied the GM-VFN scheme
to the calculation of the cross sections for $\gamma
+ \gamma \to D^{*\pm} + X$ \cite{KS1,KS2} and $\gamma p \to
D^{*\pm} + X$ \cite{KS3}. In Ref.~\cite{KS2}, we considered only the direct and
the single-resolved cross sections with $m \neq 0$. In the calculation of
the full cross section for $\gamma + \gamma \to D^{*\pm} + X$, needed for the
comparison with experimental data, {\it i.e.}\ in the sum of the direct,
single-resolved and double-resolved parts, the double-resolved
contribution was still treated in the ZM-VFN scheme with $n_f=4$.
It is the purpose of this work to apply the GM-VFN approach to the $p\bar{p}$
cross section.  The results of
this calculation can then also be applied to the cross sections of
double-resolved $\gamma \gamma$ and resolved $\gamma p$ collisions.
These cross sections play an important role due to
the partonic subprocesses $g + g \to c + \bar{c}$ and $q + \bar{q}
\to c + \bar{c}$ with charm quarks in the final state and due to
the subprocess $g + q(\bar{q}) \to c + \bar{c} + q(\bar{q})$, where $q$ is one
of the
light (massless) quarks $u$, $d$ and $s$.  These contributions and their
NLO corrections should be computed with massive charm quarks. Although
FFs for various charm mesons have been constructed from $e^{+}e^{-}$ data
\cite{ff-dother}, we shall restrict ourselves to 
inclusive $D^{*\pm}$ production and study the mass-dependent corrections
for this special final state only. Results for the inclusive production of
other charm mesons will be presented in a future publication.

Starting with $g + g \to c + \bar{c}$, the NLO corrections for
this subprocess can be split into an Abelian and two non-Abelian parts. The
Abelian part is, up to an overall constant factor, identical to the NLO
corrections to $\gamma + \gamma \to c + \bar{c}$.  For this part,
the terms in the massive theory surviving in the limit $m \to
0$, which are not present in the ZM-VFN approach, have been identified
in our earlier work \cite{KS1}. Therefore, only the two non-Abelian
parts of the NLO corrections to the gluon-gluon fusion cross section
have to be investigated in addition to the cross sections for $q + \bar{q}
\to c + \bar{c} + g$ and $g + q(\bar{q}) \to c + \bar{c} + q(\bar{q})$. 

The NLO corrections with non-zero quark mass $m$ were calculated by
several groups \cite{DEN,BKNS,BMNSS,BS}. In neither of these references,
complete formulas for the NLO corrections were published.
Fortunately, Bojak supplied us with the computer code which was
used in Ref.~\cite{BS}.  From this code, we were able to read off the complete
NLO squared matrix elements needed for the computation of the
mass-dependent cross section.  The authors of Ref.~\cite{BS} compared
their results with those of Refs.~\cite{BKNS,BMNSS} and found complete
agreement. Therefore, we use these expressions to derive the limit $m
\to 0$ and establish the subtraction terms by comparing to the
$\overline{\rm MS}$-factorized cross section derived in Ref.~\cite{ACGG}.
The latter is available to us in the form of a {\tt FORTRAN} program
\cite{BKK,KK}.
Since, in the work of
Ref.~\cite{BS}, the FFN cross section was derived with a method different from
the one used in Ref.~\cite{KS1}, namely with the phase space slicing method
for separating the infrared-divergent part from the hard part of the
cross section, we also derive the massless limit of the Abelian part
for consistency.  With this knowledge, we can compute the finite-mass
corrections for the full NLO cross section with $\overline{\rm MS}$ factorization.

The outline of our work is as follows. In Sec.~\ref{sec:bojak}, we
describe the formulae which we use to calculate the cross section for $g
+ g \to c (\bar{c}) + X$, $q + \bar{q} \to c (\bar{c}) +
X$, $g + q \to c (\bar{c}) + X$ and $g + \bar{q} \to c (\bar{c}) + X$ with
non-zero charm-quark mass.
For these cross sections, we perform the limit $m \to 0$ and
compare the results with the ZM-VFN theory of Ref.~\cite{ACGG}.  The results
are collected in Sec.~\ref{sec:zm-limit} and three appendices.  In
Sec.~\ref{sec:zm-limit}, we also present numerical results to test the
validity of the subtraction terms and show how the various terms in the
NLO cross section approach their corresponding massless limits for large
values of $p_T$.
After adding the contributions with (anti-)charm quarks in the
initial state, which are present in the ZM-VFN scheme with $n_f=4$,
as well as the contributions due to the fragmentation of gluons and light
{(anti-)quarks}, we compare our results to recent experimental data
from CDF \cite{CDF} in Sec.~\ref{sec:numresults}.  A summary and
conclusions are given in Sec.~\ref{sec:summary}.

\section{LO and NLO Differential Cross Sections}
\label{sec:bojak}

The differential inclusive cross section for the process $p+\bar{p}
\to D^{*\pm}+X$ has many contributions.  In this section, we
consider those contributions where the charm quark appears only in the
final state.  We study the charm-quark mass dependence to obtain the massless
limit, which is then compared with the ZM-VFN theory, and to
establish the influence of the $m^2/p_T^2$ terms in the GM-VFN theory
defined in the same $\overline{\rm MS}$ factorization scheme as the ZM-VFN theory.

There are only two leading-order (LO) partonic subprocesses, $g+g \to
c+\bar{c}$ and $q+\bar{q} \to c+\bar{c}$. The NLO 
corrections to these two channels comprise the virtual corrections and
gluonic bremsstrahlung contributions, $g+g \to c+\bar{c}+g$ and
$q+\bar{q} \to c+\bar{c}+g$.  In addition, the subprocesses $g+q
\to c+\bar{c}+q$ and $g+\bar{q} \to c+\bar{c}+\bar{q}$ appear
for the first time at NLO. In the following subsections, we present the
LO cross sections in order to fix the notation. Then, we explain how we
calculate the NLO corrections to $g+g \to c+\bar{c}$ and $q+\bar{q}
\to c+\bar{c}$ and the cross sections for $g+q(\bar{q}) \to
c+\bar{c}+q(\bar{q})$.

\subsection{LO Cross Section}
\label{sec:LO}

We start with the subprocess
\begin{equation}
g(k_1) + g(k_2) \to c(p_1) + \bar{c}(p_2) + [g(p_3)] \, ,
\end{equation}
where $k_1$, $k_2$ and $p_i$ ($i=1, 2, 3$) denote the four-momenta of the
incoming gluons, the outgoing charm and anti-charm quarks and
a possible gluon in the final state (in square brackets). We have the
following invariants
\begin{eqnarray}
 s &=& (k_1+k_2)^2 \, , \nonumber \\
 t_1&=& t-m^2 = (k_1-p_1)^2 - m^2 \, , \nonumber \\
 u_1 &=& u-m^2 = (k_2-p_1)^2 - m^2 \, , \nonumber\\ 
 s_2 &=& (k_1+k_2-p_1)^2 - m^2 = s+t_1+u_1 \, .
\end{eqnarray}
Here $t_1$ and $u_1$ are determined by the four-momentum of the observed
charm quark. As usual, we define the dimensionless
variables $v$ and $w$ by
\begin{equation}
 v = 1 +\frac{t_1}{s},\qquad w=-\frac{u_1}{s+t_1} \, ,
\end{equation}
so that $t_1=-s(1-v)$, $u_1=-svw$ and $s_2=sv(1-w)$. For $p_3 = 0$,
{\it i.e.}\ at LO, we have $s_2=0$ and $w=1$.

The LO cross section for $g+g \to c+\bar{c}$ is
\begin{eqnarray}
\frac{d^2\sigma _{\rm LO}^{gg}}{dvdw} = c(s) \delta (1-w)
\left(C_F-C_A\frac{t_1u_1}{s^2} \right) 
\left[ \frac{t_1}{u_1}+\frac{u_1}{t_1}+\frac{4m^2s}{t_1u_1}
\left(1-\frac{m^2s}{t_1u_1}\right) \right] \, ,
\label{eq:ggLO}
\end{eqnarray}
where 
\begin{equation}
c(s)=\frac{\pi \alpha_s ^2}{(N^2-1)s} \, ,
\end{equation} 
and all color factors have been expressed in terms of the Casimir
operators $C_F = (N^2-1)/(2N)$ and $C_A = N$, where $N$
denotes the number of colors.

The LO cross section for $q+\bar{q} \to c+\bar{c}$ reads
\begin{equation}
\frac{d^2\sigma _{\rm LO}^{q\bar{q}}}{dvdw}
= c_q(s) \delta(1-w) C_F 
\left( \frac{t_1^2+u_1^2}{s^2} + \frac{2m^2}{s} \right) \, ,
\label{eq:qqLO}
\end{equation}
where 
\begin{equation}
c_q(s)=\frac{\pi \alpha_s ^2}{Ns} \, .
\end{equation}

\subsection{NLO Cross Section}
\label{sec:NLO}

Following the notation of Ref.~\cite{BS}, the color decomposition of the NLO
squared matrix element for $g+g \to c+\bar{c}+g$ can be written
as
\begin{equation}
 |M_{gg}|^2 =g^6E_{\epsilon}^2 \frac{2}{N^2-1} 
\left[C_F^2 D_{\rm QED}+ 
\frac{1}{4} C_A^2 D_{\rm OQ} 
+ \frac{1}{4} C_A\left(C_A - 2C_F\right) 
              D_{\rm KQ} \right] \, ,
\label{colgg}
\end{equation}
where $g^2=4\pi \alpha_s$ is the strong coupling and $E_{\epsilon} =
1/(1-\epsilon)$ originates from averaging over the gluon spins in
$n=4-2\epsilon$ space-time dimensions. The squared matrix element for the
virtual corrections to $g+g \to c+\bar{c}$ can be written in a similar
fashion, there is, however, an additional contribution from quark loops which
comes
with a color factor $C_A/4$.  The Abelian contribution $D_{\rm
  QED}$ is identical to the QED part of $\gamma g \to c \bar{c} g$
in Ref.~\cite{BS1}. In addition, we have two non-Abelian parts $D_{\rm OQ}$
and $D_{\rm KQ}$. For isolating the divergences in the soft limit, Bojak
and Stratmann \cite{BS} used the same method as in Ref.~\cite{BKNS}. They slice
the $2 \to 3$ contributions into a {\it soft-gluon} and a
{\it hard-gluon} part
by introducing a small auxiliary quantity $\Delta $. In the limit
$\Delta \to 0$, the kinematics of the soft-gluon cross section is
that of the $2 \to 2$ process, so that the phase-space
integrations can be performed analytically.  After combination with the
virtual cross section, the infrared $1/\epsilon$ and the combined
infrared-collinear $1/\epsilon ^2$ singularities, all proportional to
the $n$-dimensional LO cross section, cancel. In this way, the {\it soft plus
virtual} cross section becomes finite, except for the remaining
collinear $1/\epsilon $ singularities, which cancel against the
subtraction terms in the collinear factorization procedure of the gluon
PDF of the (anti-)proton. The integration over the
phase space of the two unobserved partons in the hard part is done
analytically as far as possible, using the methods of
Refs.~\cite{BKNS,BMNSS}.

The same steps are taken to calculate the NLO corrections to the subprocess
$q+\bar{q} \to c+\bar{c}$. The squared matrix element for the real
corrections is color-decomposed in the following way
\begin{equation}
|M_{qq}|^2=g^6\frac{1}{2N}\left( C_F^2 N_{\rm QED}
+ \frac{1}{2} C_F C_A N_{\rm OK} \right) \, ,
\label{colqq}
\end{equation}
where $N_{\rm QED}$ is again the Abelian and $N_{\rm OK}$ the
non-Abelian part, which are also obtained from Ref.~\cite{BS}. A similar
decomposition is used for the virtual corrections, which receive an
additional contribution from one diagram with a quark loop in the gluon
propagator, proportional to the color factor $C_F/2$.

The subprocess
\begin{equation}
 g(k_1)+q(k_2) \to c(p_1)+\bar{c}(p_2) +q(p_3) \, ,
\label{procgq}
\end{equation}
which occurs only in NLO, has two pieces in color space. The squared
matrix element is split up according to
\begin{equation}
|M_{gq}|^2=g^6 E_{\epsilon}\frac{1}{4N}\left(C_F J_{\rm QED}+
\frac{1}{2} C_A J_{\rm OK}\right) \, .
\label{colgq}
\end{equation}
The Abelian part $J_{\rm QED}$ is, up to a factor, equal to the
corresponding squared matrix element for the subprocess $\gamma +q
\to c+\bar{c}+q$.  The squared matrix element for the crossed
subprocess $g + \bar{q} \to c + \bar{c} + \bar{q}$, {\it i.e.}\ 
Eq.~(\ref{procgq}) with the quark replaced by an anti-quark, has the same
structure as in Eq.~(\ref{colgq}), but with slightly different coefficients.

\section{Zero-Mass Limit of the Massive Cross Sections}
\label{sec:zm-limit}

In this section, we collect our results for the cross sections in the
limit $m \to 0$. We consider the following contributions: (i)
the NLO corrections to $g + g \to c + \bar{c}$, (ii) the NLO
corrections to $q + \bar{q} \to c + \bar{c}$ and (iii) the process
$g + q \to c + \bar{c} + q$ and the corresponding channel with $q
\to \bar{q}$, where $q$ denotes any of the light (massless) quarks
$u$, $d$ and $s$.  The result for the limit $m \to 0$ will in
general be different from the cross section obtained in the ZM-VFN approach,
where the mass of the charm quark is neglected from the beginning. In
the genuine ZM-VFN calculation, originally performed by Aversa {\it et al.}\ 
\cite{ACGG}, the collinear singularities connected with the charm quark
appear as $1/\epsilon$ poles in dimensional regularization.  In the
FFN theory, they appear as terms proportional to $\ln (m^2/s)$,
instead. So, in this theory, the collinear singularities are regularized
with a finite charm mass. Due to this different procedure for
regularizing the collinearly divergent contributions, different finite
terms appear. The occurrence of different finite terms in these two
regularization schemes is due to the fact that the two limits, $m
\to 0$ and $\epsilon \to 0$, are not interchangeable.
If one wants to implement the factorization of these collinearly singular
terms in the $\overline{\rm MS}$ scheme, as is done in the ZM-VFN scheme from the
start, the different finite terms, which turn up in the limit
$m \to 0$, must be subtracted. Such finite terms have already been
found for the case of the NLO corrections to $\gamma + \gamma
\to c + \bar{c}$ \cite{KS1} and $\gamma + g \to c + \bar{c}$
and for the cross section of $\gamma + q \to c + \bar{c} + q$
\cite{KS2}.

As in Refs.~\cite{KS1,KS2}, we decompose the NLO contributions
to the cross section in the limit $m \to 0$ as follows:
\begin{eqnarray}
\lim_{m\to 0}\frac{d^2\sigma_{\rm NLO}}{dvdw}
& = &
  \left(c_1 + \tilde{c}_1 \ln\frac{m^2}{s}\right) \delta(1-w)
\nonumber 
\\
& &
{}+ \left(c_2 + \tilde{c}_2 \ln\frac{m^2}{s}
  \right) \left(\frac{1}{1-w}\right)_{+}
+ c_3\left(\frac{\ln (1-w)}{1-w}\right)_{+} 
\nonumber
\\
& &
{}+ c_5 \ln v + c_6\ln (1-vw) + c_7 \ln (1-v+vw) + c_8\ln (1-v) 
\nonumber 
\\
& &
{}+ c_9 \ln w + c_{10}\ln (1-w) + c_{11} 
+ \tilde{c}_{11} \ln \frac{m^2}{s} 
\nonumber 
\\
& &
{}+ c_{12}\frac{\ln (1-v+vw)}{1-w} + c_{13}\frac{\ln w}{1-w}
+ c_{14}\frac{\ln \left(\frac{1-v}{1-vw}\right)}{1-w} \, .
\label{sigma_massless}
\end{eqnarray}
The coefficients $c_i$ are mass-independent; the dependence on the heavy-quark
mass enters only through the logarithms $\ln (m^2/s)$ and the choice
$\mu_R = \mu_F = m$. 

In the following subsections, we shall present the complete expressions for
the coefficients $c_i$. The massless limit of the cross sections from
Ref.~\cite{BS} will be compared with the results of Ref.~\cite{ACGG} obtained
in the ZM-VFN theory.

\boldmath
\subsection{Massless Limit of NLO Corrections to $g + g \to c +
  \bar{c}$}  
\label{sec:gg}
\unboldmath

Before we write down the coefficients $c_i$, we give the LO cross section
for $g + g \to c + \bar{c}$ with $m=0$. It has the simple form
\begin{equation}
\lim_{m\to 0}\frac{d^2\sigma^{g g}_{\rm LO}}{dvdw} 
= c(s) \delta(1-w) 
\tau(v)[C_F - C_Av(1-v)] \, ,
\label{sigma_LO_massless}
\end{equation}
with
\begin{equation}
\tau(v) = \frac{v}{1-v} + \frac{1-v}{v} \, .
\end{equation}
The results for the various coefficients $c_i$ are written in the form
\begin{equation}
c_i = \hat{c}_i + \Delta c_i \, ,
\end{equation}
where $\hat{c}_i$ are the results of Ref.~\cite{ACGG} in the ZM-VFN scheme and
$\Delta c_i$ are the subtraction terms needed to convert the cross section of
Ref.~\cite{BS} to the GM-VFN scheme. The coefficients are
decomposed into one Abelian part, two non-Abelian parts and a quark-loop
contribution (which only occurs for $c_1$) in the following way:
\begin{equation}
c_i = C(s) \left[C_F^2\ c^{\rm qed}_{i} 
            + \frac{1}{4} C_A^2\ c^{\rm oq}_{i} 
            + \frac{1}{4} C_A\left(C_A - 2C_F\right) 
              c^{\rm kq}_{i} 
            + \delta_{i1} \frac{C_A}{4} c_1^{\rm ql}\right] \, ,
\label{col_decomp}
\end{equation}
with
\begin{equation}
C(s) = \frac{\alpha_s^3}{2(N^2-1)s} = \frac{\alpha_s}{2\pi} c(s) \, .
\end{equation}
The expressions for the $c_i$ are lengthy and, therefore, delegated to
Appendix \ref{app:gg}.

The coefficients $\hat{c}_i$, which agree with results obtained from
Ref.~\cite{ACGG}, refer to the version where every incoming gluon is averaged
with the factor $1/[2(1-\epsilon)]$ and where the factorization of
singularities due to collinear quarks and gluons is performed in the
customary $\overline{\rm MS}$ subtraction scheme. Furthermore, deviating from
Ref.~\cite{ACGG}, in the expressions for $\hat{c}_i$ in Appendix
\ref{app:gg}, the renormalization scale $\mu_R$ and the initial- and
final-state factorization scales $\mu_F$ and $\mu_F^\prime$ are identified with
$m$, {\it i.e.}\ $\mu_R = \mu_F = \mu_F^\prime = m$.
The non-vanishing subtraction
terms in the massless limit of the FFN theory of Ref.~\cite{BS} are found
in Appendix \ref{app:gg}. They are $\Delta c_1$, $\Delta c_2$, $\Delta
c_3$, $\Delta c_5$, $\Delta c_{10}$ and $\Delta c_{11}$.  For all three
contributions $c^{\rm qed}_{i}$, $c^{\rm oq}_{i}$ and $c^{\rm kq}_{i}$, we have the relation
\begin{equation}
\Delta c_5 = \Delta c_{10} = 2 \Delta c_{11}\, .
\end{equation}
Denoting the ZM-NLO result of Ref.~\cite{ACGG} by $d^2\sigma_{\rm ZM}/dvdw$,
we can thus write
\begin{equation}
 \lim_{m \to 0} \frac{d^2\sigma_{\rm NLO}}{dvdw} = 
  \frac{d^2\sigma_{\rm ZM}}{dvdw}(\mu_R = \mu_F = \mu_F^\prime =m) 
+ \frac{d^2\sigma_{\rm sub}}{dvdw} \, ,
\end{equation}
where
\begin{eqnarray}
\frac{d^2\sigma_{\rm sub}}{dvdw} &=& 
\Delta c_1 \delta(1-w)+\Delta c_2 
\left (\frac{1}{1-w}\right )_{+}
+ \Delta c_3 \left[\frac{\ln (1-w)}{1-w}\right]_{+} 
\nonumber\\
& &
{}+\Delta c_5 \left[\ln v + \ln(1-w) + \frac{1}{2}\right] \, .
\label{subt}
\end{eqnarray}
For the first three subtraction terms proportional to $\Delta c_1$,
$\Delta c_2$ and $\Delta c_3$, one obtains simple expressions, if the
three contributions in Eq.~(\ref{col_decomp}) proportional to the color
factors $C_F^2$, $C_A^2$ and $C_A\left( C_A -
 2 C_F \right)$ are added. From the results in Appendix
\ref{app:gg}, we obtain
\begin{eqnarray}
\label{ci_first}
\Delta c_1 = (1-\ln v -\ln ^2 v) & \times & 
2 C(s) C_F \tau(v) [C_F - C_Av(1-v)]\, ,
\\
\Delta c_2 = - (2 \ln v +1 ) & \times & 
2 C(s) C_F \tau(v) [C_F - C_Av(1-v)]\, ,
\\
\label{ci_three}
\Delta c_3 = -2 & \times & 
2 C(s) C_F \tau(v) [C_F - C_Av(1-v)]\, ,
\end{eqnarray}
and
\begin{eqnarray}
\Delta c_5 = C(s) C_F 
\left( C_F \Delta c^{\rm qed}_{5} + \frac{1}{2}C_A 
\Delta c^{\rm oq}_{5} \right)\, .
\label{ci_last}
\end{eqnarray}
In the last equation, we have used $\Delta c^{\rm kq}_{5} = -\Delta c^{\rm oq}_{5}$; the
explicit expressions for these coefficients are given in Appendices
\ref{app:ggoqu} and
\ref{app:ggkqu}. Note that the last factors in
Eqs.~(\ref{ci_first})--(\ref{ci_three}) are proportional to the LO cross
section.
Finally, we have to subtract the quark-loop contribution, which
is absent in the ZM-VFN scheme, via \footnote{%
It should be noted that there is no such contribution
in the massless limit of the calculation of Ref.~\protect\cite{BKNS},
{\it i.e.}\ 
$c_1^{\rm ql}  =\Delta c_1^{\rm ql}=0$. 
On the other hand, we found non-zero coefficients
$c_1^{\rm ql}$ and $\Delta c_1^{\rm ql}$ in the massless limit
of Ref.~\protect\cite{BS} listed in
Appendix \ref{app:ggql}, which are numerically irrelevant, however.}
\begin{equation}
\Delta c_1 \to \Delta c_1 + C(s) \frac{C_A}{4} \Delta c_1^{\rm ql}
= \Delta c_1 - C(s) C_A \frac{1}{9} v (1-v)\, .
\end{equation}

The FFN theory for $g + g \to c + X$ in the limit $m
\to 0$ approaches the ZM-VFN theory 
with scales $\mu_R = \mu_F = \mu_F^\prime = m$ if the finite terms $\Delta c_1$,
$\Delta c_2$, $\Delta c_3$ and $\Delta c_5 =\Delta c_{10} = 2\Delta
c_{11}$ as given in Eqs.~(\ref{ci_first})--(\ref{ci_last}) are subtracted.
As already mentioned above, the necessity for such a subtraction is to
be expected, since the regularization of collinear singularities with a
mass parameter $m$ does not give the same result as the one with dimensional
regularization and $m=0$ from the start.

In Ref.~\cite{KS1}, it was shown that the finite subtraction terms for the
subprocess $\gamma + \gamma \to c + X$ can be obtained by a
convolution of the LO cross section with a perturbative partonic FF
$d_c^c(x,\mu)$ for the
transition from a massless to a massive
charm quark of the following form \cite{MN}:
\begin{equation}
d_c^c(x,\mu)= C_F \frac{\alpha_s}{2\pi} \left\{ \frac{1+x^2}{1-x} 
\left[\ln \frac{\mu^2}{m^2} -2\ln (1-x) -1 \right] \right\}_{+}\, .
\label{fin_state}
\end{equation}
Therefore, all Abelian terms proportional to $C_F^2$ in
Eqs.~(\ref{ci_first})--(\ref{ci_last}) can also be generated in this way.
Our explicit calculations show that also all terms proportional to
$C_F C_A$ in the equations above can be obtained as
final-state interaction contributions with $d_c^c(x,\mu)$ in
Eq.~(\ref{fin_state}) (without the term proportional to $\ln (\mu^2/m^2)$, of
course).

It is important to understand that the terms containing logarithms $\ln
(m^2/s)$, {\it i.e.}\ the coefficients $\tilde{c}_1$, $\tilde{c}_2$ and
$\tilde{c}_{11}$, have two different origins. On the one hand, these
terms are due to on-shell internal charm-quark lines in the Feynman
diagrams which become singular for $m \to 0$.
Also internal gluon and light-quark lines give
rise to singular contributions for $m \to 0$, which have to be
factorized into corresponding perturbative FFs
describing the transition from a gluon or a light quark to the heavy
charm quark. In the ZM-VFN calculation, these contributions
have been factorized as final-state singularities and are recovered by
setting $\mu_F^\prime = m$. Another part of these singular terms can be assigned
to the initial state and is found in the ZM-VFN calculation if one
sets $\mu_F = m$. On the other hand, in both the FFN and ZM-VFN calculations,
there are singularities due to internal gluon lines which
are, in both approaches, factorized as initial-state singularities.
Finally, there are logarithms due to the renormalization of $\alpha_s$.
In Appendix \ref{app:gg}, these terms are written down for the choice
$\mu_R = \mu_F = \mu_F^\prime = m$. 

For our application, choosing the scales equal to the heavy-quark mass
is not appropriate. For the case of large $p_T$ values, a common choice is
$\mu_R = \mu_F = \mu_F^\prime = \xi m_T$, where $m_T = \sqrt{m^2+p_T^2}$ is the 
transverse mass of the $D^{*\pm}$ meson and $\xi = {\cal O}(1)$.
Therefore, we have to rescale all terms proportional to $\ln m^2$.
In the following, we shall give the necessary terms for the conversion to
arbitrary scales. 

First, we present the contributions related to renormalization and
initial-state factorization of singularities due to internal on-shell
gluon lines.  These terms are present in the FFN calculation and can,
therefore, be obtained from Ref.~\cite{BS} in the limit $m \to 0$.
The rescaling is obtained by adding the following terms to the cross section
in Eq.~(\ref{sigma_massless}):
\begin{eqnarray} 
\lefteqn{
\left(\frac{d^2\sigma}{dvdw} \right)_{\rm rescal} = 
\left (\hat{d}_1 \ln \frac{\mu_R^2}{m^2} 
      + \tilde{d}_1 \ln \frac{\mu_F^2}{m^2} 
      + \tilde{\tilde{d}}_1 \ln \frac{\mu_F^{\prime2}}{m^2}
\right ) \delta(1-w)} 
\nonumber
\\
&&{}+\left( \tilde{d}_2 \ln \frac{\mu_F^2}{m^2} 
      + \tilde{\tilde{d}}_2 \ln \frac{\mu_F^{\prime2}}{m^2} \right) 
\left(\frac{1}{1-w}\right)_{+} 
+ \tilde{d}_{11} \ln \frac{\mu_F^2}{m^2} 
+ \tilde{\tilde{d}}_{11} \ln \frac{\mu_F^{\prime2}}{m^2}\, .
\label{rescal}
\end{eqnarray}
The non-zero coefficients read
\begin{eqnarray}
 \hat{d}_1 &=& C(s) \tau(v) 2 \beta_0^{(n_f - 1)} 
               \left[C_F - C_A v (1-v) \right] \, ,
\label{sub_first}
\\
 \tilde{d}_1 &=& C(s) \tau(v) \left[C_F - C_A v (1-v)\right]
                 \left\{2 C_A [\ln(1-v) - \ln(v)] - 2
                   \beta_0^{(n_f-1)}\right\} \, ,
\label{sub_firstb}
\\
 \tilde{d}_2 &=& -C(s) \tau(v) 4 C_A 
  \left[ C_F - C_A v(1-v) \right]\, ,
\label{sub_two}
\\
\tilde{d}_{11} &=& C(s) 
\left[\frac{C_A^2}{4}\ \tilde{d}_{11}^{\rm oq}
+ \frac{C_A\left(C_A - 2C_F\right)}{4}\ 
  \tilde{d}_{11}^{\rm kq}\right]\, ,
\label{sub_last}
\end{eqnarray}
with $\beta_0^{(n_f)} = 11 N /6 - n_f/3$ and
\begin{eqnarray}
\tilde{d}_{11}^{\rm oq} &=&
\frac{4 ( -1 + 5 v - 12 v^2 + 20 v^3 - 12 v^4 + 4 v^5 ) }{v_1^2 v} 
\nonumber\\
& &
{}- \frac{4 ( 1 - 4 v + 8 v^2 - 8 v^3 + 4 v^4 ) }{v_1  v w^2} 
- \frac{4 ( 3 - 9 v + 14 v^2 - 11 v^3 + 4 v^4 ) }{v_1^2 w} 
\nonumber\\
& &
{}- \frac{4 ( -1 + 5 v - 12 v^2 + 28 v^3 - 24 v^4 + 12 v^5 )  w}{v_1^2 v} 
+ \frac{16 v^3 ( 1 + v )  w^2}{v_1^2} 
\nonumber\\
& &
{}- \frac{16 v^4 w^3}{v_1^2} 
+ \frac{4 v_1  v}{X^3} 
+ \frac{4 v ( -4 + 3 v ) }{X^2} 
+ \frac{4 v ( 10 - 13 v + 5 v^2 ) }{v_1 X}\, ,
\\
\tilde{d}_{11}^{\rm kq} &=&
\frac{-4 ( -1 + 3 v - 4 v^2 + 4 v^3 ) }{v_1^2 v} 
+ \frac{4 ( 1 - 2 v + 2 v^2 ) }{v_1 v w^2} 
+ \frac{4 ( 1 - v + v^3 ) }{v_1^2 w} 
\nonumber\\
& &
{}+ \frac{4 ( -1 + 3 v - 4 v^2 + 4 v^3 )  w}{v_1^2 v} 
- \frac{4 v_1  v}{X^3} 
+ \frac{4 v_2 v }{X^2} 
- \frac{4 v ( 4 - 3 v + v^2 ) }{v_1 X}\, ,
\end{eqnarray}
where we have used the abbreviations $X = 1 - vw$ and $v_i = i -v$.  We
repeat that the expressions above are given in the limit $m \to
0$ and will be used in our ZM-VFN calculation with rescaling. In the
FFN calculation, we shall use instead the corresponding terms as given
in Ref.~\cite{BS} including their full mass dependence.

Finally, the remaining logarithms emerging in the limit $m \to 0$
due to internal charm-quark lines becoming massless are again found by
comparing the massless limit of the FFN cross section with the ZM-VFN cross
section. These terms are associated with a
rescaling of the subtraction terms in Eq.~(\ref{subt}), and, therefore, we
write them in the following form:
\begin{eqnarray} 
\lefteqn{\left(\frac{d^2\sigma}{dvdw} \right)_{\Delta \rm rescal} = 
\left (\Delta \hat{d}_1 \ln \frac{\mu_R^2}{m^2} 
      + \Delta \tilde{d}_1 \ln \frac{\mu_F^2}{m^2} 
      + \Delta \tilde{\tilde{d}}_1 \ln \frac{\mu_F^{\prime2}}{m^2}
\right ) \delta(1-w)} 
\nonumber
\\
&&{}+\left( \Delta \tilde{d}_2 \ln \frac{\mu_F^2}{m^2} 
      + \Delta \tilde{\tilde{d}}_2 \ln \frac{\mu_F^{\prime2}}{m^2} \right) 
\left(\frac{1}{1-w}\right)_{+} 
+ \Delta \tilde{d}_{11} \ln \frac{\mu_F^2}{m^2} 
+ \Delta \tilde{\tilde{d}}_{11} \ln \frac{\mu_F^{\prime2}}{m^2}\, .
\label{drescal}
\end{eqnarray}
These terms must be added to Eq.~\eqref{subt}, {\it i.e.}\ {\it subtracted}
from Eq.~(\ref{sigma_massless}), to rescale to the appropriate renormalization
and factorization scales if one wants to use the FF for the transition
$c \to D^{*+}$ and the charm PDF of the (anti-)proton.
Since, at present, we have at our disposal only PDFs and FFs, which are based
on a ZM-VFN calculation, we shall take the
corresponding coefficients for $m = 0$ as well.
Note that this does not entail any loss of accuracy, as has been discussed in
Ref.~\cite{sacot} in the context of deep-inelastic scattering.
Moreover, this fact is of great practical importance, since the known
coefficients of the ZM-VFN scheme, {\it e.g.}\ those of Ref.~\cite{ACGG}, can
simply be used, whereas their massive counterparts are unknown and can only be
obtained through a dedicated calculation.
The non-zero coefficients are given by
\begin{eqnarray} 
\Delta \hat{d}_1&=&  \frac{2}{3} C(s) \tau (v) 
               \left[C_F - C_A v(1-v)\right]\, ,
\label{d1_tilde}
\\
\Delta \tilde{d}_1&=&- \frac{2}{3} C(s) \tau (v) 
               \left[C_F - C_A v(1-v)\right]\, ,
\label{d1_tildea}
\\
\Delta \tilde{\tilde{d}}_1&=& C(s) \tau (v) C_F \left(2\ln v +
\frac{3}{2}\right) 
\left[C_F - C_A v(1-v)\right]\, ,
\label{d1_tildeb}
\\
\Delta \tilde{\tilde{d}}_2
&=& 2C(s)\tau (v)C_F 
\left[C_F - C_A v(1-v)\right]\, ,
\label{d2_tilde} 
\\
\Delta \tilde{\tilde{d}}_{11}&=&
C(s)\left[C_F^2\ \Delta \tilde{\tilde{d}}_{11}^{\rm qed} 
+ \frac{C_A^2}{4}\ 
\Delta \tilde{\tilde{d}}_{11}^{\rm oq} 
+\frac{C_A\left(C_A - 2C_F\right)}{4}\ 
\Delta \tilde{\tilde{d}}_{11}^{\rm kq} \right]\, ,
\label{d11_ttilde}
\\
\Delta \tilde{d}_{11}&=&
C(s)\left[C_F^2\ \Delta \tilde{d}_{11}^{\rm qed} 
+ \frac{C_A^2}{4}\ 
\Delta \tilde{d}_{11}^{\rm oq} 
+ \frac{C_A\left(C_A - 2C_F\right)}{4}\ 
\Delta \tilde{d}_{11}^{\rm kq} \right]\, ,
\label{d11_tilde}
\end{eqnarray}
with
\begin{eqnarray}
\Delta \tilde{\tilde{d}}_{11}^{\rm qed} &=& -\frac{v}{v_1}
+\frac{2-2v+v^2}{vw}
-\frac{v^2w}{v_1}
-\frac{2v}{Y}\, ,
\\
\Delta \tilde{\tilde{d}}_{11}^{\rm oq} &=& 
2 v (1 - 24 v )  
+ \frac{8 v_1 ( 1 - 2 v + 2 v^2 ) }{v w^2} 
+ \frac{16 ( 1 - 3 v + 2 v^2 ) }{w} 
\nonumber\\
& &
{}+ \frac{8 v^2 ( 7 - 14 v + 8 v^2 )  w}{v_1^2} 
- \frac{16 v^3 ( -1 + 2 v )  w^2}{v_1^2} 
+ \frac{16 v^4 w^3}{v_1^2} 
\nonumber\\*
& &
{}+ \frac{4 v v_1^2 }{Y^3} 
- \frac{4 v ( 6 - 11 v + 5 v^2 ) }{Y^2} 
+ \frac{2 v ( 25 - 18 v + 4 v^2 ) }{Y} \, ,
\\
\Delta \tilde{\tilde{d}}_{11}^{\rm kq} &=& 
-2 v + \frac{4 v v_1^2 }{Y^3} 
- \frac{4 v^2 v_1 }{Y^2} 
+ \frac{2 v ( 3 - 6 v + 4 v^2 ) }{Y} \, ,
\\
\Delta \tilde{d}_{11}^{\rm qed} &=& 
-\frac{2 - 2 v + 3 v^2 - 4 v^3}{v_1 v} 
+ \frac{1 + v^2}{v w} 
- \frac{( -2 + 2 v - 2 v^2 + 3 v^3 )  w}{v_1  v} 
\nonumber\\
& &
{}+ \frac{2 v v_1 }{X^3} 
- \frac{2 v}{X^2} 
- \frac{v(-3 + 4 v - 2 v^2)}{v_1 X}\, ,
\\
\Delta \tilde{d}_{11}^{\rm oq} &=& 
-\frac{2 ( 2 - v + 3 v^2 ) }{v_1^2} 
+ \frac{4 ( 1 - 2 v + 2 v^2 ) }{v_1 v w^2} 
\nonumber\\
& &
{}- \frac{2 ( -3 + 10 v - 13 v^2 + 4 v^3 ) }{v_1^2 w} 
+ \frac{4 ( 1 + v^2 )  w}{v_1^2} 
+ \frac{4 v_1 v}{X^2} 
- \frac{4 v ( -1 + 2 v ) }{X}\, ,
\\
\Delta \tilde{d}_{11}^{\rm kq} &=& -\Delta \tilde{d}_{11}^{\rm oq}\, ,
\end{eqnarray}
where $Y = 1 - v + vw$.

Comparing with the results for $\gamma +\gamma \to c+X$
\cite{KS1}, we see that the Abelian parts of Eqs.~(\ref{d1_tildeb}) and
(\ref{d2_tilde}) agree with $-\tilde{\tilde{c}}_1$ and
$-\tilde{\tilde{c}}_{2}$ in Ref.~\cite{KS1} 
(if the $C(s)$ factors are set to unity).  
Furthermore, $\Delta \tilde{\tilde{d}}_{11}^{\rm qed}$ agrees with
$-\tilde{\tilde{c}}_{11}$ and $\Delta \tilde{d}_{11}^{\rm qed}$ with
$-\tilde{c}_{11}$ in Ref.~\cite{KS1}, as one would expect.
The minus sign is due to the different convention for the logarithms
in Eq.~(42) in Ref.~\cite{KS1} as compared to Eq.~\eqref{drescal}.
We note that
$\Delta \tilde{\tilde{d}}_{11}^{\rm qed} = -\Delta c_{11}^{\rm qed}$ 
and
$\Delta \tilde{\tilde{d}}_{11}^{\rm kq} = -\Delta c_{11}^{\rm kq}$ 
(see Appendices \ref{app:ggqed} and \ref{app:ggkqu}, respectively). 
Finally, we emphasize again that the
rescaling of the $\ln m^2$ terms to arbitrary scales $\mu_R$, $\mu_F$ and
$\mu_F^\prime$ is achieved by adding Eq.~(\ref{rescal}) to 
and subtracting Eq.~(\ref{drescal}) from the
cross section in Eq.~\eqref{sigma_massless}, where all coefficients of
Eqs.~(\ref{sub_first})--(\ref{d11_tilde}) have to be taken into account.
Note also that
$\Delta \hat{d}_1$ and $\Delta \tilde{d}_1$ cancel if $\mu_R = \mu_F$.

\boldmath
\subsection{Massless Limit of NLO Corrections to $q+\bar{q} \to c
  + \bar{c}$} 
\label{sec:qq}
\unboldmath

In this section, we give the results for the massless limit of the 
hard-scattering cross sections for the NLO corrections to the subprocess
$q+\bar{q} \to c+\bar{c}$, where $q$ is any of the light quarks
assumed to be massless. The LO cross section of this process for $m=0$
is
\begin{equation}
\lim_{m \to 0} \frac{d^2\sigma _{\rm LO}^{q\bar{q}}}{dvdw} =
c_q(s) C_F \tau _q(v) \delta(1-w) \, ,
\end{equation}
where
\begin{equation}
\tau _q(v) = (1-v)^2+v^2\, .
\end{equation}
The NLO cross section is again decomposed as indicated in
Eq.~(\ref{sigma_massless}) with coefficients $c_i$ which receive
contributions from two color factors and from virtual-quark loops (which only
appear for $c_1$ and $\tilde{c}_1$).
Specifically, we have
\begin{equation}
 c_i = C_q(s) \frac{C_F}{2} 
\left(
C_F c^{\rm cf}_{i} + C_A c^{\rm ca}_{i} + \delta_{i1} c_1^{\rm ql}
\right) \, ,
\label{col}
\end{equation} 
where
\begin{equation}
C_q(s) = \frac{\alpha_s^3}{2Ns} = \frac{\alpha_s}{2\pi} c_q(s) \, .
\end{equation}
The expressions for $c^{\rm cf}_{i}$, $c^{\rm ca}_{i}$ and $c_1^{\rm ql}$ may be found in
Appendix \ref{app:qq}.
According to these results, finite subtraction terms in the massless
limit of Ref.~\cite{BS} are present in $c_1$, $c_2$, $c_3$, $c_5$, $c_{10}$
and $c_{11}$, and we have $\Delta c_5 = \Delta c_{10} = 2 \Delta c_{11}$
as in the preceding subsection.  The first three $\Delta c_i$ terms have the
following simple form:
\begin{eqnarray}
\Delta c_1 = (1-\ln v -\ln ^2 v) & \times & 2 C_q(s) \tau_q(v) C_F^2 \, ,
\label{Del_first}
\\
 \Delta c_2 = - (1+2 \ln v) & \times & 2 C_q(s) \tau_q(v) C_F^2 \, ,
\\
 \Delta c_3 = -2 & \times & 2 C_q(s) \tau_q(v) C_F^2 \, ,
\end{eqnarray}
and
\begin{eqnarray}
 \Delta c_5 = 2 C_q(s) C_F^2 \left(
v - \frac{2 v v_1^2 }{Y^3} + \frac{2 v^2 v_1}{Y^2} 
- \frac{3 v - 6 v^2 + 4 v^3}{Y} \right) \, .
\label{Del_last}
\end{eqnarray}
The complete subtraction contribution for the $q\bar{q}$ cross section has
the form of Eq.~\eqref{subt} with the $\Delta c_i$ terms given in
Eqs.~(\ref{Del_first})--(\ref{Del_last}) and Appendix \ref{app:qq}. Again,
the subtraction terms agree with the convolution of the perturbative
FF of Eq.~(\ref{fin_state}) with the LO $q\bar{q}$ cross
section.
We remark that the $\Delta c_i$ terms only appear in the QED part.

As in the $gg$ channel, the results in Appendix \ref{app:qq} are given
for the choice $\mu_R = \mu_F = \mu_F^\prime = m$.
For our application, we need
the cross section for arbitrary scales.  To make the transition, we use
Eq.~(\ref{rescal}), where the non-zero contributions to the
coefficients $\hat{d}_i$, $\tilde{d}_i$ and $\tilde{\tilde{d_i}}$ have
the following form:
\begin{eqnarray}
\label{qscal_first}
 \hat{d}_1 &=& C_q(s) \frac{C_F}{2} \tau_q(v) 4 \beta_0^{(n_f-1)}
\, ,
\\
\label{qscal_firstb}
 \tilde{d}_1 &=& - C_q(s) \tau_q(v) C^2_{\rm F} [3 + 2 \ln v - 2 \ln(1-v)]\, ,
\\
\label{qscal_two}
\tilde{d}_2 &=& -4 C_q(s) \tau_q(v) C_F^2  \, ,
\\
\tilde{d}_{11} &=& C_q(s) C_F^2 \biggl[ 
\frac{1-8v+12v^2-6v^3}{v_1} 
-\frac{1-2v+2v^2}{w} 
\nonumber\\
& &
{}+\frac{2 v_2 v^2 w}{v_1}  
-\frac{2v^3w^2}{v_1} +\frac{v}{X} \biggr]\, .
\label{qscal_first1}
\end{eqnarray}
Of course, for the calculation of the FFN cross section, the
corresponding mass-dependent contributions are used as given in
Ref.~\cite{BS}.  These terms rescale all the $\ln m^2$ contributions due to
the renormalization and the factorization of internal light-quark lines.
Therefore, the shift from renormalization due to internal quark loops is
proportional to $n_f-1$.

To eliminate all $\ln m^2$ terms, we must also take into account the
contribution from charm-quark loops and the one from internal charm-quark
lines that is factorized into the FF.  These
terms are taken into account by using Eq.~(\ref{drescal}), and the
corresponding coefficients are given by
\begin{eqnarray}
\label{qscal_last1}
\Delta \hat{d}_1 &=&  C_q(s) \tau_q(v) \frac{2}{3} C_F \, ,
\\
\label{qscal_last1b}
\Delta \tilde{\tilde{d}}_1 &=& C_q(s) \tau_q(v) 
\frac{C_F^2}{2} (3+4\ln v)\, ,
\\
\label{qscal_last2}
\Delta \tilde{\tilde{d}}_2 &=& C_q(s) \tau_q(v) 2 C_F^2 \, ,
\\
\Delta \tilde{\tilde{d}}_{11}&=&
C_q(s) \left(\frac{C_F^2}{2} 
  \Delta \tilde{\tilde{d}}_{11}^{\rm cf} 
  + \frac{C_F C_A}{2} 
  \Delta \tilde{\tilde{d}}_{11}^{\rm ca} 
  \right)\, ,
\label{qscal_last}
\end{eqnarray}
where
\begin{eqnarray}
  \Delta \tilde{\tilde{d}}_{11}^{\rm cf} &=& 
\frac{2 v ( 2 - 3 v + 2 v^2 ) }{v_1} 
+ \frac{2 ( 1 - 2 v + 2 v^2 ) }{w} 
- \frac{4 v^3 w}{v_1} 
+ \frac{4 v^3 w^2}{v_1} 
\nonumber\\
& &
{}+ \frac{4 v v_1^2 }{Y^3} 
- \frac{4 v^2 v_1 }{Y^2} 
+ \frac{2 v ( 1 - 6 v + 4 v^2 ) }{Y}\, ,
\\
  \Delta \tilde{\tilde{d}}_{11}^{\rm ca} &=& 
4 v_2  v - 4 v^2 w 
- \frac{4 v v_1^2}{Y^3} 
+ \frac{4 v ( 3 - 5 v + 2 v^2 ) }{Y^2} 
- \frac{2 v ( 9 - 12 v + 4 v^2 ) }{Y}\, .
\end{eqnarray}
In the general case, the contributions written down in
Eqs.~(\ref{qscal_first})--(\ref{qscal_first1}) have to be added
and the contributions in
Eqs.~(\ref{qscal_last1})--(\ref{qscal_last}) have to be subtracted.

\boldmath
\subsection{Massless Limit of $g+q \to c+\bar{c}+q$ and $g + \bar{q} \to c
  + \bar{c} + \bar{q}$} 
\label{sec:gq}
\unboldmath

The processes $g+q \to c+\bar{c}+q$ and $g+\bar{q} \to c+\bar{c}+\bar{q}$ appear
for the first time at NLO.  In the massless limit, the corresponding
cross sections are decomposed as explained in Eq.~\eqref{sigma_massless}.
Since there is no LO part, one has $c_1 = c_2 = c_3 =0$.  The remaining
coefficients are decomposed with respect to the color factors $C_{\rm
  F}$ and $C_A$ as
\begin{equation}
\label{eq:gq-col}
c_i = C_q(s) \left(C_F c^{\rm cf}_{i} 
      + C_A c^{\rm ca}_{i} \right) \, .
\end{equation}
The resulting coefficients $c_i^{\rm cf}$ and $c_i^{\rm ca}$ are given
in Appendix \ref{app:gq} for the case that the observed transverse
momentum is due to the charm quark. If the anti-charm quark is observed
with a given value of $p_T$, the coefficients are different. The differences
$c_i^{\bar{c} - c} = c_i(\bar{c}~\mbox{observed}) -
c_i(c~\mbox{observed})$ are listed in Appendix \ref{app:gqbar}. Charge
conjugation invariance relates the cross sections for quarks and
anti-quarks in the initial state as $d\sigma/dp_T (g + q \to
c_{\rm observed} + \bar{c} + q) = d\sigma/dp_T (g + \bar{q} \to
\bar{c}_{\rm observed} + c + \bar{q})$.

We find that, for all $gq$ and $g\bar{q}$ channels, there are no extra
finite subtraction terms, {\it i.e.}\ we have $\Delta c_i = 0$ in all cases.
Comparing the coefficients
$c_i^{\rm cf}$ and $c_i^{{\rm cf}, \bar{c} - c}$ with
the results for $\gamma + q \to c + \bar{c} + q$ given in
Ref.~\cite{KS2}, we find agreement, except for $\Delta c_{11}$, which was 
found to be non-zero for $\gamma + q \to c + \bar{c} + q$.
The subtraction terms in Ref.~\cite{KS2} were obtained by matching the FFN
result of Ref.~\cite{MCG} with the ZM-VFN result of Ref.~\cite{LG}.
In order to clarify this mismatch, we compare the formula for
$\gamma + q \to c + \bar{c} + q$ of Ref.~\cite{LG} with the one from
Ref.~\cite{aur}, which we extracted from the {\tt FORTRAN} program used in
that paper, to find that the two disagree.
On the other hand, the result in Ref.~\cite{aur} agrees with the Abelian part
of the result for $g + q \to c + \bar{c} + q$ in Ref.~\cite{ACGG}.
Relying on Ref.~\cite{aur}, we then conclude that we also have
$\Delta c_{11}=0$ for $\gamma + q \to c + \bar{c} + q$ in Ref.~\cite{KS2}.

As in the previous cases, the formulas written down in Appendix
\ref{app:gq} are for the scale choice $\mu_R = \mu_F = \mu_R^\prime = m$.
The
transition to arbitrary scales $\mu_R$, $\mu_F$ and $\mu_F^\prime$ is
obtained, as
above, using Eq.~\eqref{rescal} with the appropriate coefficients. In
the present case, $\tilde{d}_1$ and $\tilde{d}_2$ do not contribute, only
$\tilde{d}_{11}$ is non-zero. If we factor out $C_q(s)$, $C_F$ and
$C_A$ as in Eq.~\eqref{eq:gq-col}, we obtain the coefficients
$\tilde{d}^{\rm cf}_{i}$ and $\tilde{d}^{\rm ca}_{i}$, which correspond to the initial-state factorization
of singularities due to internal gluons, in the following form:
\begin{eqnarray}
\label{gqscal_1}
\tilde{d}^{\rm cf}_{11} &=& 
\frac{1 - 4 v + 9 v^2 - 6 v^3 + 2 v^4}{v_1^2} 
+ \frac{-1 + 2 v - 4 v^2 + 3 v^3 - v^4}{v_1^2 w} 
\nonumber\\
& &
{}- \frac{( 1 - 4 v + 9 v^2 - 6 v^3 + 2 v^4 )w}{v_1^2} 
- \frac{v}{2 X^2} 
+ \frac{v(3 - v)}{2 v_1 X}\, ,
\\
\tilde{d}^{\rm ca}_{11} &=& 
\frac{v^2}{v_1^2} 
- \frac{v^2 ( 2 + v^2 )  w}{v_1^2} 
+ \frac{2 v^3 ( 1 + v )  w^2}{v_1^2} 
- \frac{2 v^4 w^3}{v_1^2} 
+ \frac{v}{2 X}\, .
\label{gqscal_2}
\end{eqnarray}

In addition, we need the rescaling terms associated with initial-state
singularities of internal charm-quark lines and of final-state
singularities of gluon lines splitting into $c\bar{c}$ pairs. These
contributions vanish for the choice $\mu_F = \mu_F^\prime = m$, which is used in
Appendix \ref{app:gq}. To convert the cross sections to arbitrary
factorization scales $\mu_F$ and $\mu_F^\prime$, we have to take into account the
additional rescaling terms denoted as $\Delta \tilde{d}_i$ and
$\Delta \tilde{\tilde{d}}_i$,
\begin{eqnarray}
\label{gqscal_3}
\Delta \tilde{d}^{\rm cf}_{11} &=& 
\frac{1+v^2}{2 v_1^2 w} (1-2 w + 2 w^2)\, ,
\\
\label{gqscal_4a}
\Delta \tilde{\tilde{d}}^{\rm cf}_{11} &=&
-2 v^2 (1-w) 
+ \frac{1-2 v + 2 v^2}{2 w}
- \frac{v_1 v}{2 Y^2}
+ \frac{v (3 - 2 v)}{2 Y} \, ,
\\
\Delta \tilde{\tilde{d}}^{\rm ca}_{11} &=& 
-v^2 + \frac{v^2 ( 2 - 4 v + 3 v^2 )  w}{v_1^2} 
- \frac{2 v^3 ( -1 + 2 v )  w^2}{v_1^2} 
+ \frac{2 v^4 w^3}{v_1^2} 
+ \frac{v}{2 Y}\, .
\label{gqscal_4}
\end{eqnarray}

\subsection{Numerical Test of Subtraction Terms and Study of
  Mass-Dependent Corrections}
\label{sec:numtest}

The calculation of the subtraction terms, in particular those in
Subsections \ref{sec:gg} and \ref{sec:qq}, was rather involved. Special
care had to be exercised in order to recover all the terms proportional
to $\delta(1-w)$, $1/(1-w)_+$, $[\ln(1-w)/(1-w)]_+$ and the remaining
terms in the decomposition of Eq.~(\ref{sigma_massless}). In particular, the
delta-function terms were difficult to calculate, since they received
contributions from several places, the virtual corrections and both the soft
and hard parts of the real corrections.
Some of these contributions contained a dependence on the slicing
parameter. The cancellation of these contributions in the final results
provided a partial check of our analytical calculations, in particular
for the coefficients of the two plus-distributions. 

In order to check that the $\Delta c_i$ terms presented in Subsections
\ref{sec:gg} and \ref{sec:qq} are correct, and also to see how the
various contributions to $d^2\sigma^{ab}/dvdw$ (where $ab$ stands for
the channels $gg$, $q\bar{q}$, $gq$ and $g\bar{q}$) written
down in Subsection \ref{sec:NLO} behave as functions of the transverse
momentum $p_T$ of the $D^{*\pm}$ meson, we calculate the NLO
corrections in four different ways:
\begin{itemize}
\item[1)] Using the results of Ref.~\cite{BS}, we
  calculate the cross sections for the channels $gg$ ($g + g \to
  c + X$), $q\bar{q}$ ($q + \bar{q} \to c + X$) and $gq$ ($g +
  q(\bar{q}) \to c + X$) in the FFN scheme. The corresponding results
  will be labeled by $\sigma_m$ (massive calculation).
\item[2)] We calculate the same cross sections in the limit $m
  \to 0$, using Eq.~(\ref{sigma_massless}) with the corresponding
  coefficients as given in Subsection \ref{sec:zm-limit} and the
  appendices. Notice that $m$ is kept at its physical value whenever it
  appears logarithmically in Eq.~(\ref{sigma_massless}).
  The results will be denoted $\sigma_0$ (massless limit of the massive
  calculation).
\end{itemize} 
For both the massive calculation and its massless limit, we shall then
consider the effect of the subtraction terms described by the
coefficients $\Delta c_i$:
\begin{itemize}
\item[3)] The subtracted massive calculation, denoted by
  $\sigma^{\Delta}_m$, is obtained by subtracting Eq.~(\ref{subt}) with the
  corresponding coefficients written down in the appendices from
  $\sigma_m$. This corresponds to the GM-VFN scheme.
\item[4)] The subtracted massless cross sections are calculated by
  subtracting Eq.~(\ref{subt}) from $\sigma_0$ and will be denoted by
  $\sigma^{\Delta}_0$. This prescription is identical to the ZM-VFN scheme.
\end{itemize}
In all cases, we start from a calculation using $\mu_R = \mu_F = \mu_F^\prime = m$
and take into account the rescaling to other renormalization and
factorization scales in two steps:
\begin{itemize}
\item Firstly, we rescale to $\mu_R = \mu_F = \mu_F^\prime = 2m_T$ adding the
  terms due to renormalization and initial-state factorization of
  singularities related to internal gluon lines, according to
  Eq.~(\ref{rescal}).
  This scale choice is to prevent the value of $\mu_F^\prime$ from falling below the
  starting scale $2m$ for the $\mu_F^\prime$ evolution of the FF for low values of
$p_T$.
\item Finally, we subtract the remaining rescaling terms of Eq.~(\ref{drescal})
  related to internal charm-quark lines which become singular in the
  limit $m \to 0$.
\end{itemize}

In the following, we always normalize the cross sections to the LO ZM-VFN
cross sections calculated from Eqs.~\eqref{eq:ggLO}
and \eqref{eq:qqLO}, {\it i.e.}\ we consider the
cross section ratios $R_{ab}$ given by
\begin{equation}
R_{ab} = 
\frac{d\sigma^{ab}/dp_T}{d\sigma^{ab}_{\rm LO}/dp_T (m = 0)} \, .
\label{tratio}
\end{equation}
For the $gq$ and $g\bar{q}$ channels, we normalize to the LO ZM-VFN
$gg$ cross section, since, for the $gq$ channel, there is no LO contribution.
All results are given at the hadron level, {\it i.e.}\ the partonic cross
sections are convoluted with the (anti-)proton PDFs and the FF for
$c \to D^{*+}$.
We average over $D^{*+}$ and $D^{*-}$ mesons.
We use the PDF set CTEQ6M \cite{CTEQ6M} and the FF set obtained in 
Ref.~\cite{BKK} by fitting OPAL data at NLO.
Although the CTEQ6M PDFs were determined in the ZM-VFN scheme with
$n_f=5$, for the time being, we only include three light quark flavors
along with the gluon as incoming partons for all values of $\mu_F$.
Since, at this point, we wish to focus on effects related to the
hard-scattering cross sections, the LO cross sections are evaluated using the
same conventions concerning $\alpha_s(\mu_R)$, proton PDFs and $D^{*\pm}$ FF
as in NLO.
We consider $d\sigma/dp_T$ at $\sqrt{S}=1.96$~TeV as a
function of $p_T$ with $y$ integrated over the range $-1.0 < y <1.0$.
As for the QCD input parameters, we take the charm-quark mass to be
$m = 1.5$~GeV
and evaluate $\alpha_s^{(n_f)}(\mu_R)$ with $n_f = 4$ and asymptotic scale
parameter $\Lambda_{\rm QCD}^{(4)} = 328$~MeV, which corresponds to
$\alpha_s^{(5)}(m_Z) = 0.1181$. The renormalization scale $\mu_R$
and the factorization scales $\mu_F$ and $\mu_F^\prime$ are set equal,
$\mu_R = \mu_F = \mu_F^\prime$.  Details for the calculation of
$d\sigma / dp_T$ from
$d^2\sigma / dv dw$ have been given for direct $\gamma \gamma$ scattering in
Eq.~(45) of Ref.~\cite{KS1}. In this equation, the photon distribution 
functions must be replaced by the (anti-)proton PDFs of the gluon and
the light (anti-)quarks.

%
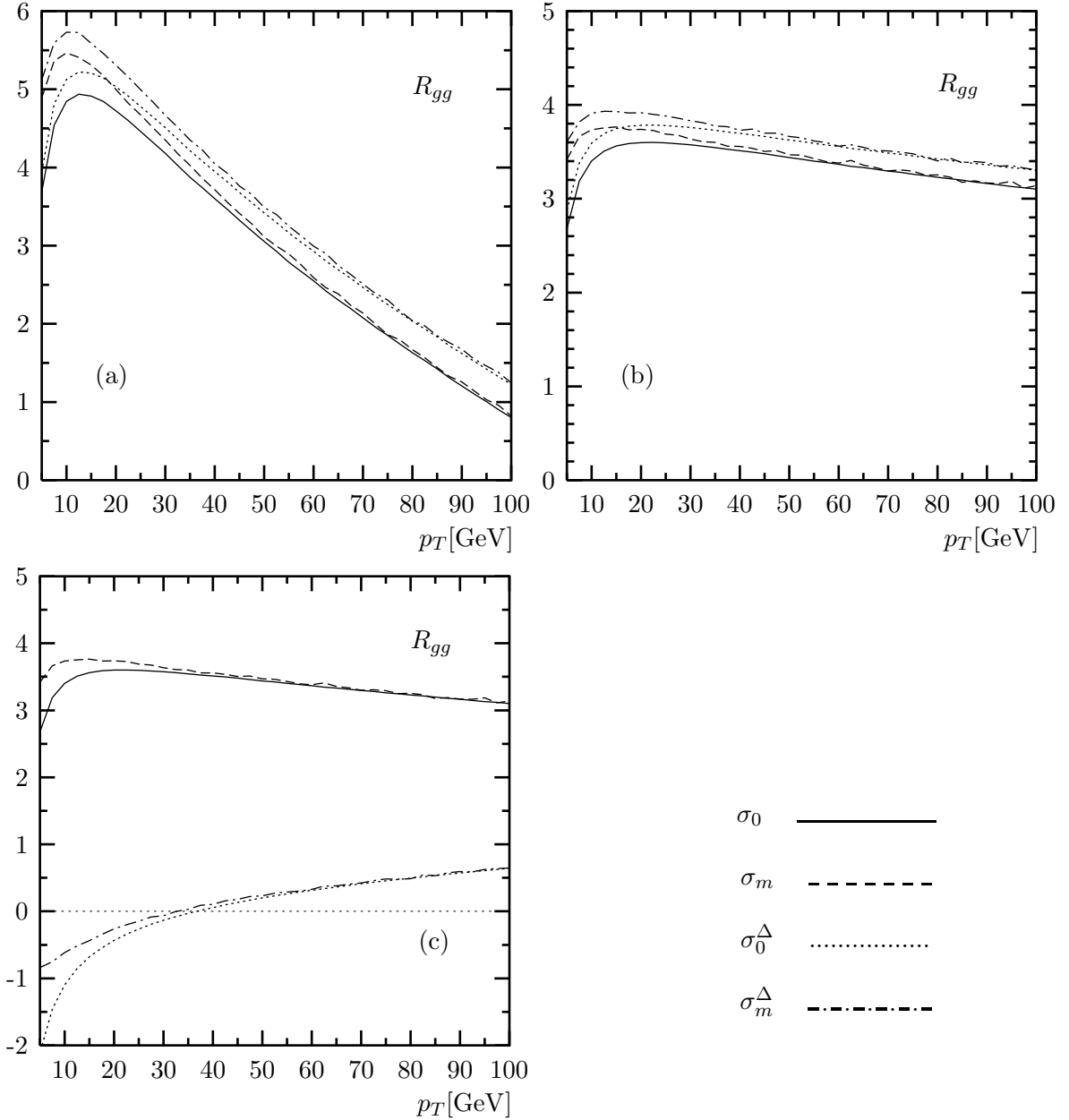
\begin{figure}[ht!] 
\unitlength 1mm
\begin{picture}(158,160)
\put(0,78){\begin{minipage}[b][70mm][b]{70mm}
\include{fig1}
\end{minipage}}
\put(15,103){(a)}
\put(78,78){\begin{minipage}[b][70mm][b]{70mm}
\include{fig2}
\end{minipage}}
\put(93,103){(b)}
\put(-2,-6){\begin{minipage}[b][70mm][b]{70mm}
\include{fig3}
\end{minipage}}
\put(63,19){(c)}
\put(90,10){\begin{minipage}[b][70mm][b]{70mm}
\thicklines
$\sigma_0$ ~~ \line(1,0){20.5}\\[1ex]
$\sigma_m$ ~~ \line(1,0){1.5}~\line(1,0){1.5}~\line(1,0){1.5}~%
\line(1,0){1.5}~\line(1,0){1.5}~\line(1,0){1.5}~\line(1,0){1.5}
\\[1ex]
$\sigma^{\Delta}_0$ ~~ $.................$\\[1ex]
$\sigma^{\Delta}_m$ ~~ 
\line(1,0){1.5}\hspace{0.4mm}\raisebox{-0.2mm}{$.$}\hspace{0.4mm}%
\line(1,0){1.5}\hspace{0.4mm}\raisebox{-0.2mm}{$.$}\hspace{0.4mm}%
\line(1,0){1.5}\hspace{0.4mm}\raisebox{-0.2mm}{$.$}\hspace{0.4mm}%
\line(1,0){1.5}\hspace{0.4mm}\raisebox{-0.2mm}{$.$}\hspace{0.4mm}%
\line(1,0){1.5}\hspace{0.4mm}\raisebox{-0.2mm}{$.$}\hspace{0.4mm}%
\line(1,0){1.5}%
\end{minipage}}
\end{picture}
\caption{$gg$ contribution to $p\bar{p} \to D^{*\pm} + X$
  including Abelian and non-Abelian parts, normalized to the LO cross
  section with $m=0$. (a): Renormalization and factorization scales are
  $\mu_R = \mu_F = \mu_F^\prime = m$ (but fixed at $2.1m$ in $\alpha_s$,
  PDFs and FF).  (b): Same as in part (a), but for
  $\mu_R = \mu_F = \mu_F^\prime = 2m_T$
  including rescaling due to renormalization and initial-state
  factorization of singularities originating from internal gluons. (c): Same
  as in part
  (a), but including full rescaling to $\mu_R = \mu_F = \mu_F^\prime = 2m_T$.}
\label{fig:fig1}
\end{figure}

The results for the $gg$ channel are shown in Fig.~\ref{fig:fig1}.  In
Fig.~\ref{fig:fig1}a, the renormalization and factorization scales are
$\mu_R = \mu_F = \mu_F^\prime = m$, but the scales in $\alpha_s$, the PDFs and
the FF are fixed at $2.1m$, to stay above the starting scale of
the FF.  The full line in Fig.~\ref{fig:fig1}a corresponds to the massless
limit of the FFN calculation as
derived in Sec.~\ref{sec:gg} ($\sigma_0$), and the dashed curve is the
result of the massive calculation ($\sigma_m$).  We see that the massive
cross section approaches the massless one very slowly at large
values of $p_T$.  At $p_T=20$~GeV, the difference between the massive and
massless cross sections is still of the order of 6\%. The ratio
$R_{gg}$ for the massive cross section is always larger than its
massless limit in the $p_T$ range between 5 and 100~GeV.  From
this comparison, we conclude that, in the $gg$ channel, the terms
proportional to $m^2/p_T^2$ are particularly large and lead to an
increase of the massive cross section as compared to the massless
approximation. Similar observations were made in Ref.~\cite{CGN}, where the
massive and massless calculations were compared as functions of the mass
$m$ for fixed value of $p_T$. For our application, we are interested in the
massive and massless cross sections, where the finite $\Delta c_i$ terms
derived in Sec.~\ref{sec:gg} are subtracted.  This leads to the
dashed-dotted ($\sigma_m^{\Delta}$) and the dotted ($\sigma_0^{\Delta}$)
curves in Fig.\ \ref{fig:fig1}a.  We have checked that our result for
the massless calculation after subtraction, {\it i.e.}\ the dotted curve, is
in perfect agreement with the results in the ZM-VFN scheme obtained using
the {\tt FORTRAN} program of Refs.~\cite{BKK,KK}. This comparison demonstrates
that the finite $\Delta c_i$ terms, which, if subtracted, should produce
the ZM-VFN cross section, are correct.
Their subtraction from the FFN result
will give the GM-VFN result, which approaches the ZM-VFN result at large
values of $p_T$.  We see
that the contribution of these finite terms is by no means negligible.

The ratios plotted in Fig.~\ref{fig:fig1}a show that, in the low- to
medium-$p_T$ range, the NLO cross section is up to a factor of about
five larger than the LO one (note that the numerator of
$R_{gg}$ is the sum of the LO result and the NLO corrections).
This is not surprising,
since we have chosen very low values for the renormalization and
factorization scales.  Since we are interested in the region where $p_T
\agt  m$, a better choice of scales is $\mu \approx p_T$. As usual, we choose
$\mu_R = \mu_F = \mu_F^\prime = 2m_T$, which can be used for both small and
large values of $p_T$. To obtain the cross sections at this scale, we must add
the terms
proportional to $\ln(m^2/\mu^2)$ as described in Sec.~\ref{sec:gg} in the
massless limit and the corresponding terms for the FFN scheme
contained in Ref.~\cite{BS}. In Fig.~\ref{fig:fig1}b, we first show the
results for the cross section ratios including the rescaling due to
renormalization and initial-state factorization, {\it i.e.}\ adding
Eq.~\eqref{rescal} using the coefficients of
Eqs.~\eqref{sub_first}--\eqref{sub_last} for the massless calculation and the
corresponding mass-dependent terms of Ref.~\cite{BS} for the massive
calculation. As in Fig.~\ref{fig:fig1}a, we show four curves corresponding to
the massive and massless calculations, in either case without and with finite
terms subtracted.
In this case, the cross section ratios exhibit a much weaker $p_T$ dependence.
The QCD correction ($K$)
factor is somewhat smaller now, but it is still large, showing that the
perturbative expansion for the $gg$ channel is not converging very well. We
observe that the massive cross sections converge to the corresponding
massless cross sections with increasing value of $p_T$ as in
Fig.~\ref{fig:fig1}a.
The effect of the subtraction of the finite terms is
slightly smaller, since the added $\ln (\mu^2/m^2)$ terms apparently
have smaller $m^2/p_T^2$ corrections.  The curves for the massive theory
lie always above the massless approximation, as in Fig.~\ref{fig:fig1}a.

If it were not for the choice $n_f=4$ in $\alpha_s$ and the fact that
the CTEQ6M proton PDFs are evolved according to the ZM-VFN scheme, so that the
charm and bottom PDFs participate in the DGLAP evolution for sufficiently
high values of $\mu_F$, the
unsubtracted result $\sigma_m$ in Fig.~\ref{fig:fig1}b would
correspond to the cross section in the genuine FFN scheme.
We have seen that this theory is characterized by large NLO
corrections.  This has its origin in the fact that the contributions of
the would-be collinear divergences related to incoming and outgoing
charm-quark lines are not yet subtracted, {\it i.e.}\ they are still left at
the factorization scales $\mu_F=\mu_F^\prime = m$.
It is clear that these contributions
must be also evaluated at the factorization scales $\mu_F=\mu_F^\prime= 2m_T$,
since
our FF for $c \to D^{*+}$ is evolved to this scale and we want to
include the contributions from the charm content of the incoming
(anti-)proton.  To do this, we must include the additional contributions
proportional to $\ln (\mu_R^2/m^2)$,
$\ln (\mu_F^2/m^2)$ and $\ln (\mu_F^{\prime2}/m^2)$ in
Eqs.~\eqref{d1_tilde}--\eqref{d11_tilde}.  The result is shown in
Fig.~\ref{fig:fig1}c.
The subtracted massive calculation ($\sigma_m^{\Delta}$) is
represented by the dash-dotted curve and the result of the
subtracted massless calculation ($\sigma_0^{\Delta}$) by the dotted one.
The latter curve again agrees with the ZM-VFN result calculated on the basis
of Ref.~\cite{ACGG}.  For comparison, we also show the results for
$\sigma_0$ and $\sigma_m$ from Fig.~\ref{fig:fig1}b before the additional
logarithmic terms from Eqs.~\eqref{d1_tilde}--\eqref{d11_tilde} are
subtracted, and also without subtraction of the finite $\Delta c_i$ terms. The
lower two curves are our final results for the $gg$ channel.  The cross
section ratios are negative for $p_T \alt 30$~GeV and rise up to approximately
0.6 at $p_T = 100$~GeV.  The massive cross section approaches its
massless limit with increasing value of $p_T$. As we shall see later, the
$gg$ channel
is only important at small values of $p_T$; at higher values of $p_T$, the
total
cross section is dominated by the contribution due to the charm content of the
(anti-)proton. Since this dominating contribution does not contain
finite-$m$ corrections, the effect of mass-dependent terms in the
$gg$ channel will be suppressed in the final result.

%
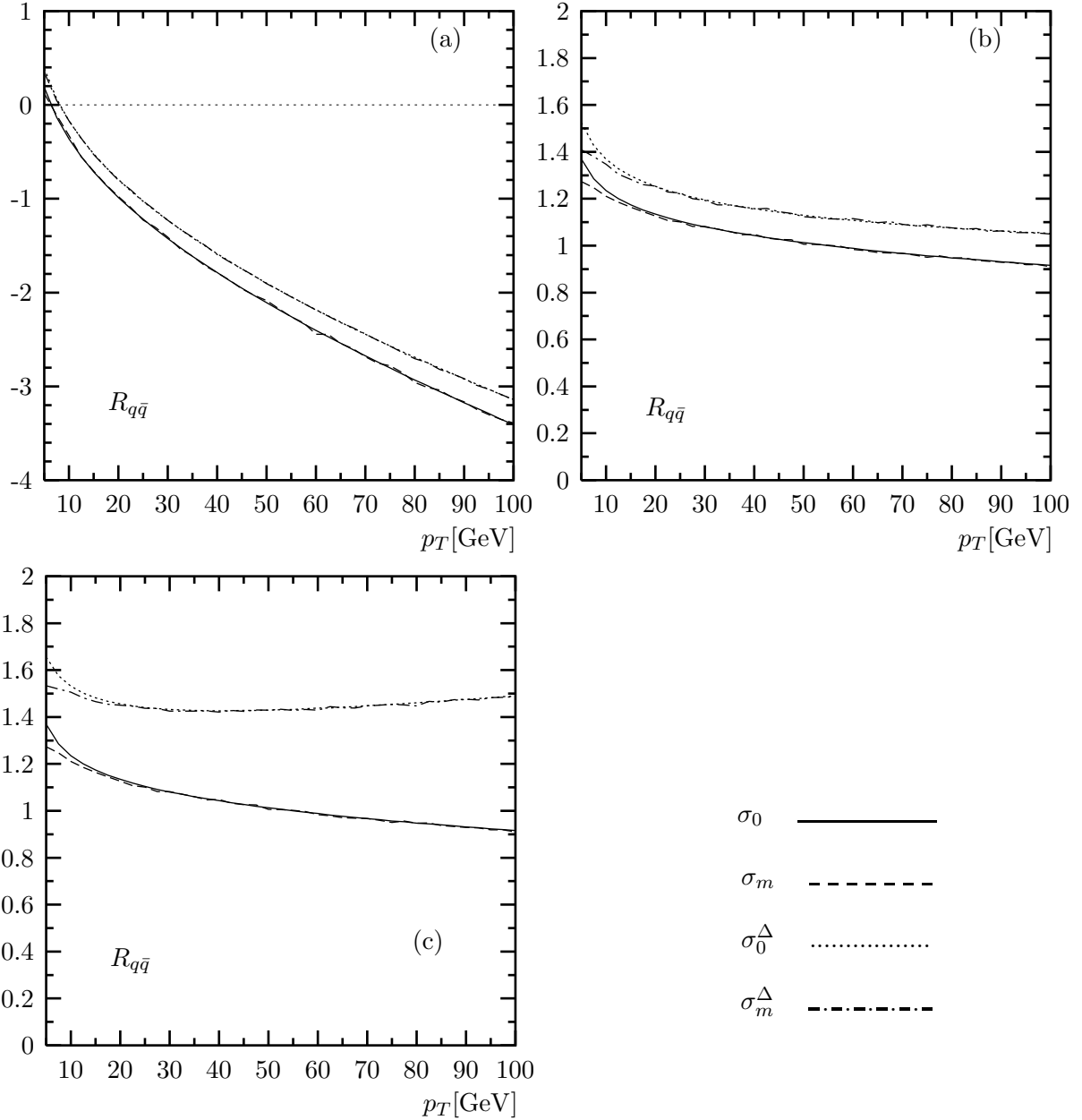
\begin{figure}[t!] 
\unitlength 1mm
\begin{picture}(158,160)
\put(-1.5,78){\begin{minipage}[b][70mm][b]{70mm}
\include{fig4}
\end{minipage}}
\put(64.5,153){(a)}
\put(76.5,78){\begin{minipage}[b][70mm][b]{70mm}
\include{fig5}
\end{minipage}}
\put(144.5,153){(b)}
\put(-3,-6){\begin{minipage}[b][70mm][b]{70mm}
\include{fig6}
\end{minipage}}
\put(62,19){(c)}
\put(90,10){\begin{minipage}[b][70mm][b]{70mm}
\thicklines
$\sigma_0$ ~~ \line(1,0){20.5}\\[1ex]
$\sigma_m$ ~~ \line(1,0){1.5}~\line(1,0){1.5}~\line(1,0){1.5}~%
\line(1,0){1.5}~\line(1,0){1.5}~\line(1,0){1.5}~\line(1,0){1.5}
\\[1ex]
$\sigma^{\Delta}_0$ ~~ $.................$\\[1ex]
$\sigma^{\Delta}_m$ ~~ 
\line(1,0){1.5}\hspace{0.4mm}\raisebox{-0.2mm}{$.$}\hspace{0.4mm}%
\line(1,0){1.5}\hspace{0.4mm}\raisebox{-0.2mm}{$.$}\hspace{0.4mm}%
\line(1,0){1.5}\hspace{0.4mm}\raisebox{-0.2mm}{$.$}\hspace{0.4mm}%
\line(1,0){1.5}\hspace{0.4mm}\raisebox{-0.2mm}{$.$}\hspace{0.4mm}%
\line(1,0){1.5}\hspace{0.4mm}\raisebox{-0.2mm}{$.$}\hspace{0.4mm}%
\line(1,0){1.5}%
\end{minipage}}
\end{picture}
\caption{$p(q)+\bar{p}(\bar{q})$ contribution to $p\bar{p} \to
  D^{*\pm} + X$ including Abelian and non-Abelian parts, normalized to
  the LO cross section with $m=0$. (a): Renormalization and factorization
 scales are $\mu_R = \mu_F = \mu_F^\prime = m$ (but fixed at
  $2.1m$ in $\alpha_s$, PDFs and FF). (b): Same as in part (a), but for
  $\mu_R = \mu_F = \mu_F^\prime = 2m_T$
  including rescaling due to renormalization and
  initial-state factorization. (c): Same as in part (a), but including full
  rescaling to $\mu_R = \mu_F = \mu_F^\prime = 2m_T$.}
\label{fig:fig3}
\end{figure}

In Fig.~\ref{fig:fig3}, we present results for the $q\bar{q}$ channel.
The cross section ratios shown here are normalized to the corresponding
LO cross section for the $q\bar{q}$ initial states. Since this
  normalization differs from the LO $gg$ cross section, the ratios in
  Figs.~\ref{fig:fig1} and \ref{fig:fig3} should not be added.
In Fig.~\ref{fig:fig3}a, the ratios with the renormalization and
factorization scales equal to $m$ are shown. The massless and massive
cross sections, {\it i.e.}\ the full ($\sigma_0$) and the dashed ($\sigma_m$)
curves with no subtraction of finite terms and the dotted
($\sigma_0^{\Delta}$) and dashed-dotted ($\sigma_m^{\Delta}$) curves
with $\Delta c_i$ terms subtracted, almost coincide. This means that, for the
$q\bar{q}$ channel, the $m^2/p_T^2$ terms are negligibly small in the
considered $p_T$ range. We notice that the NLO cross section for this
channel is negative, except for small values of $p_T$.
The subtraction terms are
non-negligible over the whole $p_T$ range.  In Fig.~\ref{fig:fig3}b, the
cross section ratio at the scale $\mu_R = \mu_F = \mu_F^\prime = 2m_T$
including the rescaling due to renormalization and initial-state factorization
according to Eqs.~(\ref{rescal}), (\ref{qscal_first})--(\ref{qscal_first1})
is shown.  At this scale, the cross section is positive for all $p_T$
values. The influence of the $m^2/p_T^2$ terms is somewhat larger now
due to additional mass-dependent contributions in the terms
proportional to $\ln (\mu^2/m^2)$. Also the difference
due to the subtraction of the finite $\Delta c_i$ terms is larger, since
the PDFs and the FF are evaluated at much larger scales. We note that the
massive ratio is smaller than the massless one. If the logarithmic
rescaling terms due to internal charm-quark lines in Eqs.~(\ref{drescal}),
(\ref{qscal_last1})--(\ref{qscal_last}) are subtracted, we obtain the cross
section ratios presented in Fig.~\ref{fig:fig3}c.  These results are
needed in our final analysis.  The dashed-dotted
($\sigma_m^{\Delta}$) and dotted ($\sigma_0^{\Delta}$) curves
lie very near to each other showing that the $m^2/p_T^2$ terms are much
smaller in this channel.  Also in this case, we checked that our result
for the massless calculation with subtraction and rescaling included
agrees perfectly with the ZM-VFN result of Ref.~\cite{ACGG}. For
comparison, the unsubtracted massive ($\sigma_m$) and massless ($\sigma_0$)
results from Fig.~\ref{fig:fig3}b are again shown in
Fig.~\ref{fig:fig3}c.  In contrast to the $gg$ channel, the last rescaling due
to singular internal charm-quark lines leads to an increased ratio.
             
%
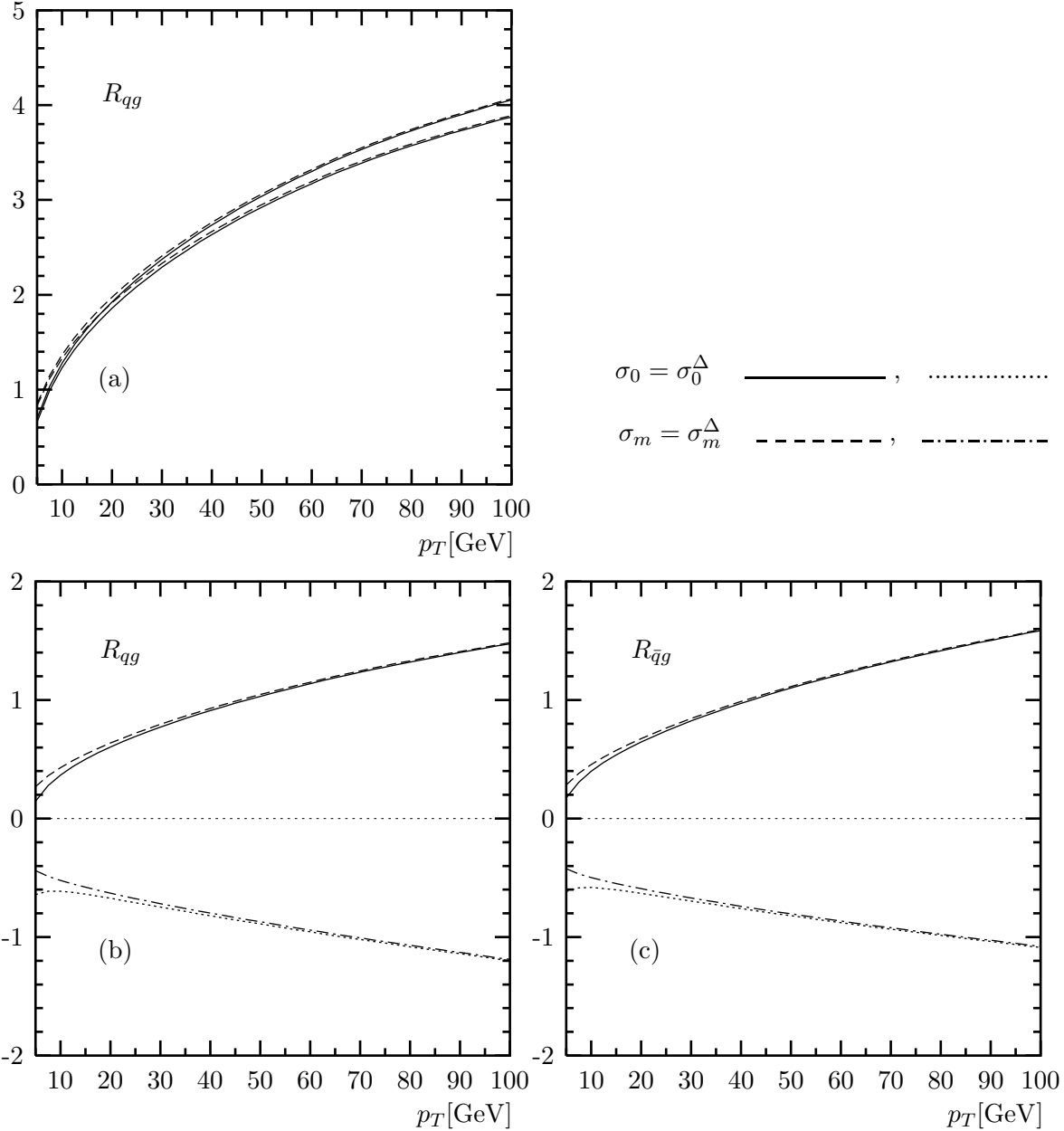
\begin{figure}[t!] 
\unitlength 1mm
\begin{picture}(158,160)
\put(1,78){\begin{minipage}[b][70mm][b]{70mm}
\include{fig7}
\end{minipage}}
\put(17,103){(a)}
\put(90,95){\begin{minipage}[b][70mm][b]{70mm}
\thicklines
$\sigma_0 = \sigma^{\Delta}_0$ ~~ \line(1,0){20.5}~, 
~~$.................$
\\[1ex]
$\sigma_m = \sigma^{\Delta}_m$ 
~~ \line(1,0){1.5}~\line(1,0){1.5}~\line(1,0){1.5}~%
\line(1,0){1.5}~\line(1,0){1.5}~\line(1,0){1.5}~\line(1,0){1.5}
~,~~ 
\line(1,0){1.5}\hspace{0.4mm}\raisebox{-0.2mm}{$.$}\hspace{0.4mm}%
\line(1,0){1.5}\hspace{0.4mm}\raisebox{-0.2mm}{$.$}\hspace{0.4mm}%
\line(1,0){1.5}\hspace{0.4mm}\raisebox{-0.2mm}{$.$}\hspace{0.4mm}%
\line(1,0){1.5}\hspace{0.4mm}\raisebox{-0.2mm}{$.$}\hspace{0.4mm}%
\line(1,0){1.5}\hspace{0.4mm}\raisebox{-0.2mm}{$.$}\hspace{0.4mm}%
\line(1,0){1.5}%
\end{minipage}}
\put(-1,-6){\begin{minipage}[b][70mm][b]{70mm}
\include{fig8}
\end{minipage}}
\put(17,19){(b)}
\put(77,-6){\begin{minipage}[b][70mm][b]{70mm}
\include{fig9}
\end{minipage}}
\put(95,19){(c)}
\end{picture}
\caption{(a): $p(q)+\bar{p}(g)$ contribution to $p\bar{p} \to
  D^{*\pm} + X$ including Abelian and non-Abelian parts, normalized to
  the LO cross section of the $gg$ channel with $m=0$. Renormalization and
  factorization scales are $\mu_R = \mu_F = \mu_F^\prime = m$ (but fixed at
  $2.1m$ in $\alpha_s$, PDFs and FF). Lower curves: observed charm,
  upper curves: observed anti-charm. (b): Same as in part (a) for
  observed charm, but with renormalization and factorization
  scales chosen as $\mu_R = \mu_F = \mu_F^\prime = 2m_T$. (c): Same as in part
  (b), but for observed anti-charm.}
\label{fig:fig5}
\end{figure}

Finally, we discuss the $m^2/p_T^2$ contributions for the $gq$ and $g\bar{q}$
channels. For definiteness,
we consider the $p(q)+\bar{p}(g)$ contributions to $p + \bar{p} \to
D^{*\pm}+X$ separately for the cases where the $D^{*\pm}$ meson originates
from a charm or an anti-charm quark.
We normalize the cross sections to the LO $gg$-channel
cross section. Using as scales $m$ in the same way as in Fig.~\ref{fig:fig1}a,
we obtain the full and dashed curves in Fig.~\ref{fig:fig5}a corresponding to
massless ($\sigma_0$) and massive ($\sigma_m$) cross sections, respectively.
The upper curves are for observed anti-charm and the lower ones
for observed charm. These two sets of curves differ only little. The $m^2$
power corrections are small over the whole $p_T$ range.  If we rescale
to the scale $2m_T$ using Eqs.~(\ref{rescal}),
(\ref{gqscal_1})--(\ref{gqscal_2}), we obtain the massless (full curves) and
massive (dashed curves) results for observed charm in Fig.~\ref{fig:fig5}b and
those for observed anti-charm in Fig.\ \ref{fig:fig5}c.
Here, the ratio of cross sections is positive.  We observe that the
massive cross section is slightly larger than the massless one. If we
include the full rescaling of logarithmic terms due to charm or anti-charm
exchange using Eqs.~(\ref{rescal}) and (\ref{drescal}) with
Eqs.~(\ref{gqscal_1})--(\ref{gqscal_4}), we obtain the lower curves in
Fig.~\ref{fig:fig5}b and \ref{fig:fig5}c, respectively, for the massive
(dashed-dotted curves) and massless (dotted curves) cross sections.  
The latter ones are again in perfect agreement with the ZM-VFN results of
Ref.~\cite{ACGG}.
These results will be used for the final cross section.

%
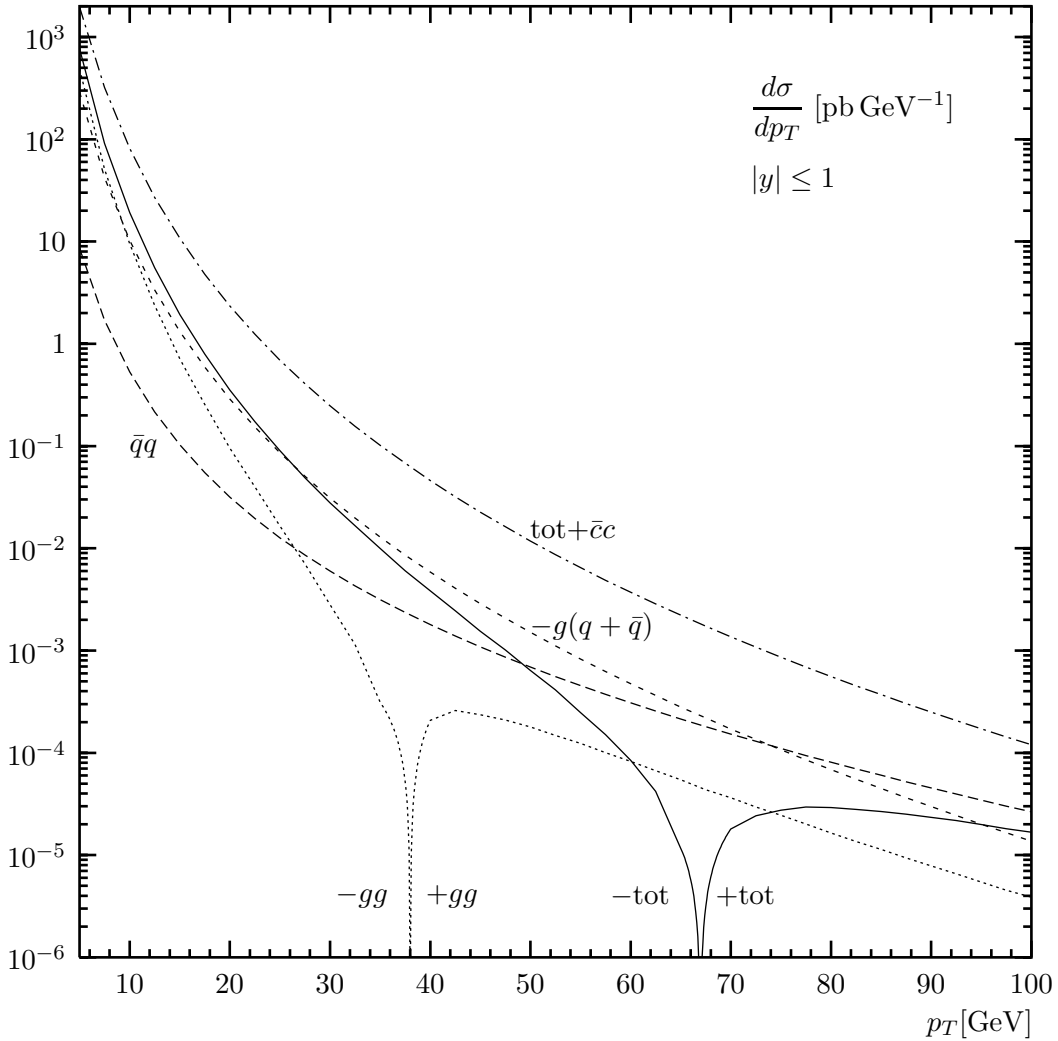
\begin{figure}[ht!] 
\unitlength 1mm
\begin{picture}(158,150)
\put(5,0){\begin{minipage}[b][70mm][b]{70mm}
\include{fig11}
\end{minipage}}
\end{picture}
\caption{Partial cross sections for $p\bar{p} \to D^{*\pm}+X$ in
  the ZM-VFN scheme. Renormalization and factorization
  scales are $\mu_R = \mu_F = \mu_F^\prime = 2m_T$. The sum of the $gg$,
  $q\bar{q}$, $gq$ and $g\bar{q}$ channels is labeled {\it tot}, the total cross
  section including incoming (anti-)charm quarks {\it tot${}+c\bar{c}$}.}
\label{fig:fig7}
\end{figure}

\begin{figure}[ht!] 
\unitlength 1mm
\begin{picture}(158,150)
\put(5,0){\begin{minipage}[b][70mm][b]{70mm}
\include{fig13}
\end{minipage}}
\end{picture}
\caption{Partial cross sections normalized to the result for all
  contributions including the $c\bar{c}$ initial state. Lower and upper
  curves correspond to the massless and massive calculations, respectively. }
\label{fig:fig9}
\end{figure}
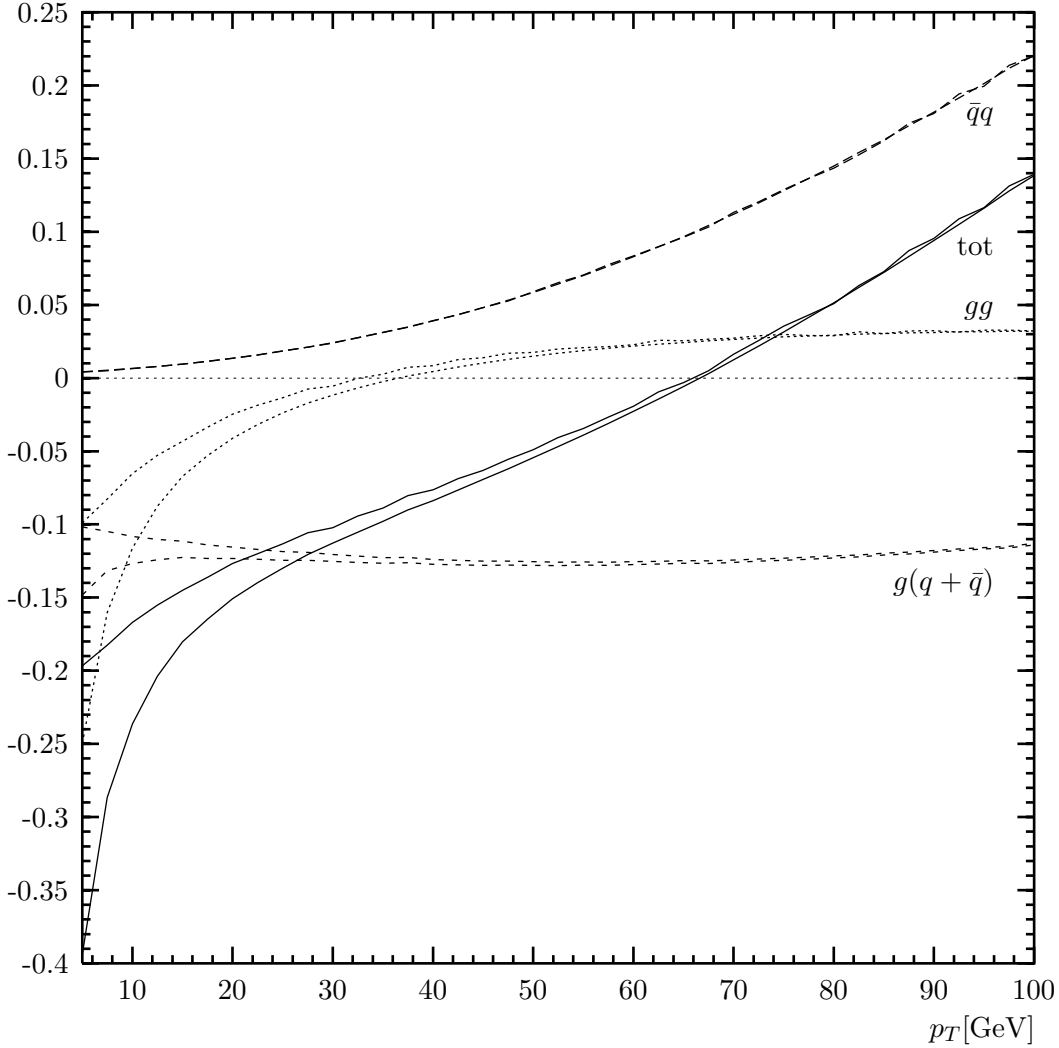

So far, we have only considered the channels with gluon or light
(anti-)quarks in the initial state and (anti-)charm quarks in the final state. To
these contributions, we must add the cross sections with (anti-)charm quarks
in the initial state. Such results will be shown in the next three
figures.  From Figs.~\ref{fig:fig1}c, \ref{fig:fig5}b and c, we see that
the $gg$ and $gq$ channels yield negative contributions (for $p_T
< 30$~GeV in the $gg$ case and for all values of $p_T$ in the $gq$ case),
while only the $q\bar{q}$ channel gives a positive contribution everywhere.
Their sum is still negative for $p_T < 65$~GeV,
as can be seen in Fig.~\ref{fig:fig7}, where we have plotted the sum of
these three channels as the solid curve labeled {\it tot}. Only
when the contributions from incoming (anti-)charm quarks are added, we
obtain, as a
meaningful result, a positive cross section, which is shown in
Fig.~\ref{fig:fig7} as the dashed-dotted curve labeled {\it tot${}+c\bar{c}$}.
These results are for the subtracted massless approach ($\sigma_0^\Delta$) and
represent the cross section for
inclusive $D^{*\pm}$ production, including only the contribution due to the
fragmentation of a (anti-)charm quark, but not the ones due the
to fragmentation of a gluon or a light (anti-)quark. In addition to
these results, we also show in Fig.~\ref{fig:fig7} the partial cross sections
for the $gg$, $q\bar{q}$ and $g(q+\bar{q})$ channels. The latter channel
contains all contributions with a gluon and a light (anti-)quark
coming from the (anti-)proton. We see that the sum of all
components yields a cross section with a smooth, steeply falling $p_T$
dependence.
The comparison with the equivalent results with massive (anti-)charm
quarks is made in Fig.~\ref{fig:fig9}, where we show for each partial
result the ratio with respect to the full result including the
contribution from $c\bar{c}$ initial states.  The upper and lower curves
correspond to the subtracted massive ($\sigma_m^\Delta$) and massless
($\sigma_0^\Delta$) calculations,
respectively.  As we can see, the $m^2/p_T^2$ corrections in the sum of
all contributions with gluons or light (anti-)quarks in the initial state
labeled {\it tot} are modest, except at small values of $p_T$.
At the smallest $p_T$ value considered, $p_T =5$~GeV, the
massless cross section is only increased by roughly 20\% when these 
corrections are included, although
the contribution labeled {\it tot} is substantially increased, from
$-773$~pb\,GeV$^{-1}$
for the massless calculation to $-386$~pb\,GeV$^{-1}$ for the massive one.
We emphasize that all results in Figs.~\ref{fig:fig7} and \ref{fig:fig9}
refer to the scale choice $\mu_R = \mu_F = \mu_F^\prime = 2m_T$.

\section{Comparison with CDF Data}
\label{sec:numresults}

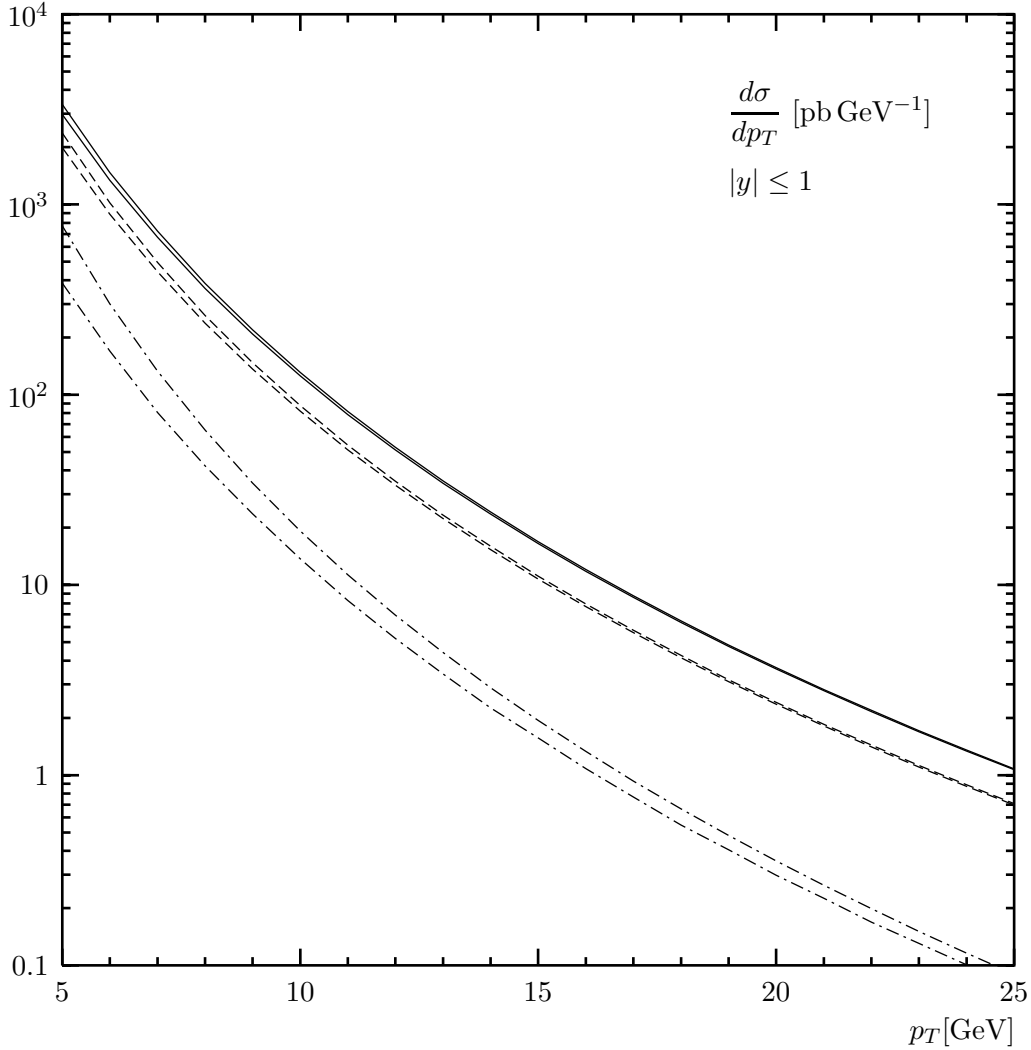
\begin{figure}[t!] 
\unitlength 1mm
\begin{picture}(158,150)
\put(5,0){\begin{minipage}[b][70mm][b]{70mm}
\include{fig14}
\end{minipage}}
\end{picture}
\caption{$p_T$ spectrum for $p\bar{p} \to D^{*\pm}+X$ including
  all contributions. Renormalization and factorization scales are
 $\mu_R = \mu_F = \mu_F^\prime = 2m_T$. Dash-dotted lines: the (negative!)
  contributions with gluons and light (anti-)quarks in the initial
  state; dashed lines: including (anti-)charm quarks in the initial state;
  full lines: including the contribution from $g, q, \bar{q} \to
  D^{*\pm}$ fragmentation. Upper and lower curves correspond to the
  massive and massless calculations, respectively.}
\label{fig:fig10}
\end{figure}

Before we can compare our final results with experimental data, we have
to add another contribution, which was not yet discussed, but is
non-negligible for the experimental situation at the Tevatron: in fact, the
observed $D^{*\pm}$ meson may also be produced through the fragmentation
of a gluon or a light (anti-)quark. Appropriate
$g, q, \bar{q} \to D^{*\pm}$ FFs are contained in the OPAL set of
Ref.~\cite{BKK}, where they are generated via NLO evolution assuming that
they vanish at the starting scale.
In the ZM-VFN scheme, gluon and light (anti-)quark fragmentation already 
contributes at LO.
By contrast, these types of contribution do not exist in the FFN scheme,
where the collinear $g\to c\bar{c}$ splitting is treated perturbatively
instead, starting only in NLO.
Since the GM-VFN scheme is to be constructed in such a way that it merges
with the ZM-VFN scheme at large values of $p_T$, it is clear that the
gluon and light (anti-)quark fragmentation contributions must be accommodated
in the GM-VFN framework as well.
A certain part of these contributions carries mass dependence, because of
internal charm-quark lines and external ones that do not lead to
fragmentation.
In the FFN scheme, the analogous mass dependence only comes in beyond NLO.
Therefore, we ignore this mass dependence for the time being and adopt the
gluon and light (anti-)quark fragmentation contributions from the ZM-VFN 
analysis \cite{ACGG}.
These contributions amount to slightly more than $30\%$,
almost independent of $p_T$, as can be seen in Fig.~\ref{fig:fig10}.
The bulk is due to gluon fragmentation.
For photoproduction in $ep$ and $\gamma \gamma$ collisions, this
contribution was found to be negligible \cite{KS1,KS2,KS3}.

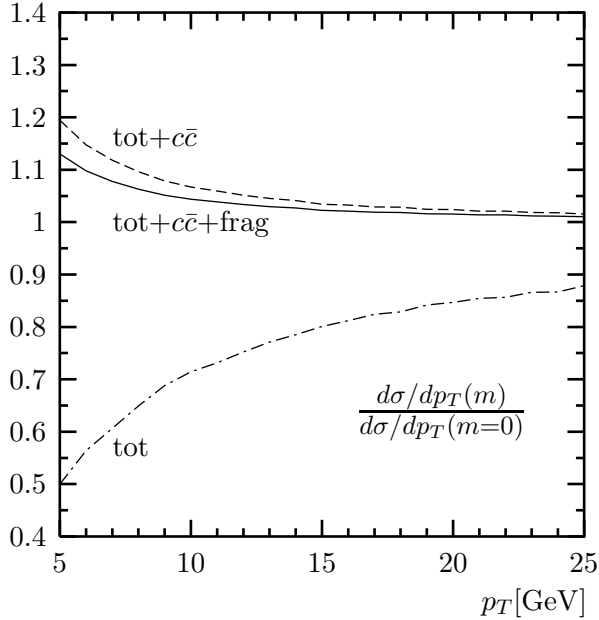
\begin{figure}[t!] 
\unitlength 1mm
\begin{picture}(158,85)
\put(38,-5){\begin{minipage}[b][70mm][b]{70mm}
\include{fig15}
\end{minipage}}
\put(88,21){\large$\frac{d\sigma/dp_T(m)}{d\sigma/dp_T(m=0)}$}
\end{picture}
\caption{Ratios of the subtracted massive and massless differential
  cross sections $d\sigma/p_T$ for $p\bar{p} \to D^{*\pm}+X$ with
  $|y| \le 1$ including (a) all contributions (full line), (b) all
  contributions with incoming gluons, light (anti-)quarks and (anti-)charm
  quarks, but without the $g, q, \bar{q} \to D^{*\pm}$ fragmentation
  contributions (dashed line), and (c) the sum of the contributions with
  only gluons and light\break
 (anti-)quarks in the initial state (dash-dotted
  line). Renormalization and factorization scales are 
  $\mu_R = \mu_F = \mu_F^\prime = 2m_T$.  }
\label{fig:fig11}
\end{figure}

The effect of mass-dependent terms is very much reduced in the final
cross section, since the parts which have to be calculated with $m=0$
are large. Therefore, one cannot expect that mass terms would increase
the theoretical predictions towards cross sections as high as 
observed in the CDF experiment \cite{CDF} . The size of the mass-dependent
terms is visualized in Fig.~\ref{fig:fig11}, where we show the ratios of cross
sections calculated with $m \ne 0$ to those with $m=0$. For
the (negative) contributions due to incoming gluons and light (anti-)quarks,
mass-dependent terms lead to a reduction in size by $50\%$
at $p_T = 5$ GeV. At this value of $p_T$, the ratio of massive over
massless results is reduced to $1.19$ and $1.13$ when the contributions
from incoming (anti-)charm quarks and from $g \to D^{*\pm}$
fragmentation, respectively, are included.

\begin{figure}[t!] 
\unitlength 1mm
\begin{picture}(158,150)
\put(5,0){\begin{minipage}[b][70mm][b]{70mm}
\include{fig16}
\end{minipage}}
\end{picture}
\caption{Variation of the $p_T$ spectrum of $p\bar{p} \to
  D^{*\pm}+X$ with the renormalization and factorization scales in the
  GM-VFN scheme. The central solid curve is for
  $\mu_R = \mu_F = \mu_F^\prime = m_T$, the upper and lower dashed curves
  represent
  the maximum and minimum cross sections found by varying $\mu_R$, $\mu_F$
  and $\mu_F^\prime$
  independently within a factor of 2 up and down relative to the
  central values.  The contributions from (anti-)charm quarks in the initial
  state and from $g, q, \bar{q} \to D^{*\pm}$ fragmentation are
  included. The data points are from CDF \cite{CDF} for the average of
  $D^{*+}$ and $D^{*-}$ production. }
\label{fig:fig12}
\end{figure}
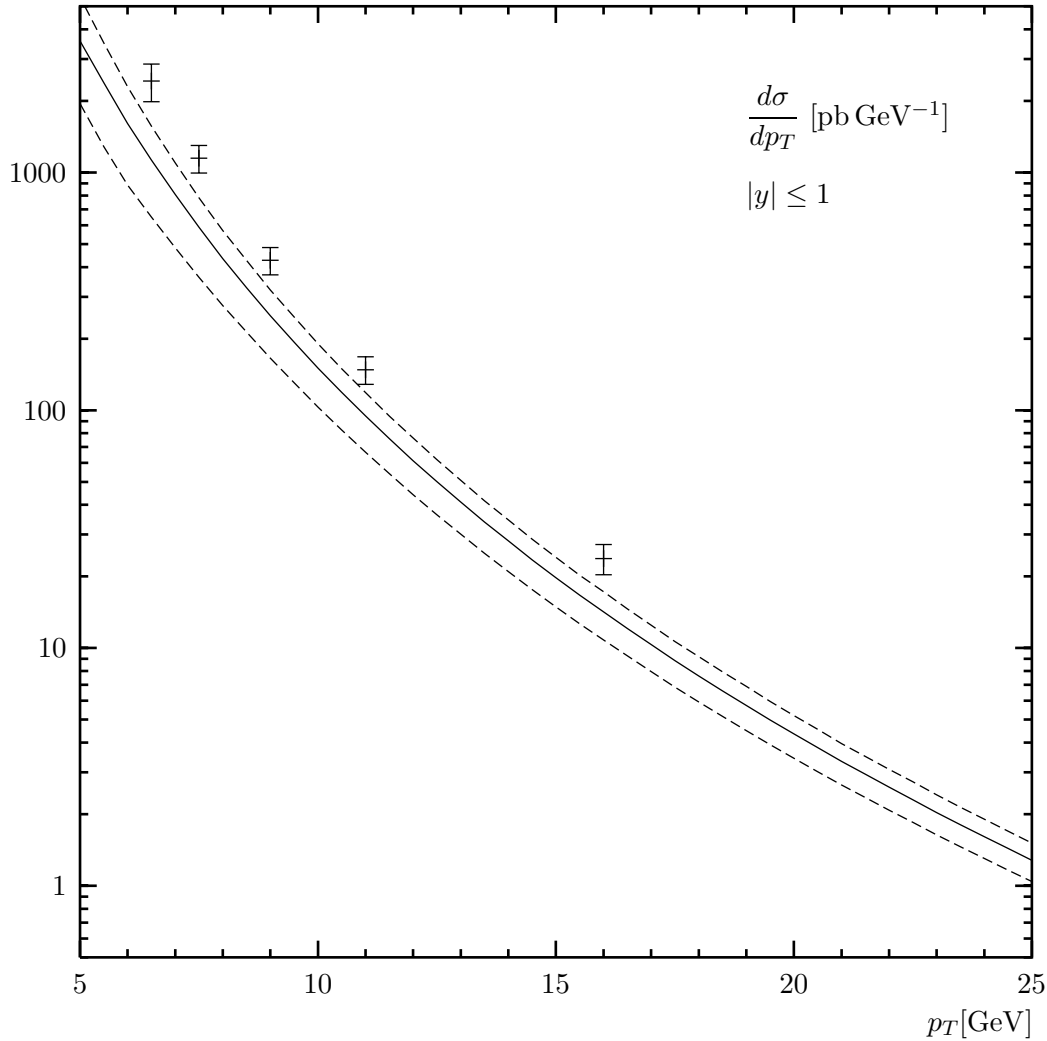

Since the effect of mass-dependent terms is very much reduced in the
final cross section, it is clear that a variation of the value of the
charm-quark mass does not contribute much to the theoretical uncertainty.
Whereas the sum of the contributions due to gluons and light (anti-)quarks in
the initial state varies by $-20.5\%$ ($+17.8\%$) at $p_T = 5$~GeV if
$m$ is changed from our default value of 1.5 GeV to 1.2 (1.8)~GeV, the
cross section also including those terms which can be calculated in the
massless approach only ({\it i.e.}\ those with\break
(anti-)charm quarks in the
initial state and from $g, q, \bar{q} \to D^{*\pm}$ fragmentation) varies by
only $-2.5\%$ ($+1.8\%$). The theoretical uncertainty is, therefore,
dominated by the choice of the renormalization and factorization scales.
The results presented in Fig.~\ref{fig:fig10} were obtained for $\mu_R =
\mu_F = \mu_F^\prime = 2m_T$.
A conservative mode of scale variation frequently encountered in the
literature is to independently vary the values of $\mu_R/m_T$, $\mu_F/m_T$ and
$\mu_F^\prime/m_T$ between 1/2 and 2 about the default value 1.
We will adopt this prescription for the comparison with experimental data from
the CDF Collaboration \cite{CDF}.
This leads to
large changes at $p_T = 5$ GeV (for the massless calculation in the
range between $-41\%$ and $+30\%$, and for the massive calculation
between $-46\%$ and $+56\%$). At $p_T = 25$ GeV, the variations are
smaller and cover the range $\pm 20\%$. Minimal values for
$d\sigma/dp_T$ are obtained for maximal values of $\mu_R$ and $\mu_F^\prime$ and
minimal value of $\mu_F$, and maximum values are reached for minimal values of
$\mu_R$ and $\mu_F^\prime$ and maximal value of $\mu_F$.
These results are presented in Fig.~\ref{fig:fig12} as a band
of predicted values.
A comparison with experimental data from CDF
\cite{CDF}, also shown in this figure, shows that the data prefer a small
renormalization scale and a large
initial-state factorization scale. However, even for the choice $\mu_R =
m_T/2$ and $\mu_F = 2m_T$, the theoretical results in the GM-VFN scheme
still undershoot the measured cross sections by $20\%$ to $34\%$.

Residual sources of theoretical uncertainty include the variations in the 
available PDF and FF sets.
The new generation of NLO proton PDF sets exhibit only minor differences.
Similarly, the two NLO FF sets that were determined in Ref.~\cite{BKK} by
fitting slightly incompatible ALEPH and OPAL data yield rather similar
predictions, the difference being of order 10\% or less in the kinematic
range of the CDF data \cite{CDF}.
The inclusion of these additional errors would broaden the theoretical error
band in Fig.~\ref{fig:fig12} only insignificantly.

We stress that our results do not include a contribution due to
bottom-quark production. This part can be identified in the experimental
analysis and was, in fact, removed from the CDF data shown in
Fig.~\ref{fig:fig12}. 

\section{Summary and Conclusions}
\label{sec:summary}

In this work, we calculated the NLO corrections to
inclusive charm production in $p\bar{p}$ collisions in two approaches
using massive or massless charm quarks. By deriving the massless limit
from the massive theory (FFN scheme), we could show that this limit differs
from the
genuine massless version with $\overline{\rm MS}$ factorization (ZM-VFN scheme) by finite
corrections. We adjusted subtraction terms and thus managed to
establish a massive theory with $\overline{\rm MS}$ subtraction (GM-VFN scheme) which
approaches the massless theory (ZM-VFN scheme) with increasing transverse
momentum. With these
results and including the contributions where a (anti-)charm quark occurs as
an incoming parton and those where a gluon or light (anti-)quark fragments,
we calculated the inclusive
$D^{*\pm}$ cross section in $p \bar{p}$ collisions at $\sqrt{S} = 1.96$~TeV
using realistic non-perturbative FFs, which are subject to proper DGLAP
evolution \cite{dglap} and manifestly universal \cite{Coll}.
Our central prediction somewhat undershoots a recent measurement by the CDF
Collaboration \cite{CDF}, but reasonable agreement can be
reached by adjusting the renormalization and factorization scales within a
plausible range of tolerance.

We made the observation that, in contrast to other experimental
situations such as $\gamma\gamma$ or $\gamma p$ collisions, the
contribution from $g \to D^{*\pm}$ fragmentation is large.
The FFs which we used here had been obtained from
fits to data from the CERN Large Electron-Positron Collider (LEP).
One may speculate, therefore,
whether these data leave enough room to adjust the $g \to
D^{*\pm}$ FF and improve the agreement with the Tevatron data. We
plan to come back to this question in a future publication. 

\begin{acknowledgments}
We are grateful to I. Bojak for providing his NLO {\tt FORTRAN} code
for the hadroproduction of heavy flavors and for clarifying comments, and to
S. Kretzer for useful discussions.
The work of I. Schienbein was supported by DESY.
This work was supported by the Bundesministerium f\"ur Bildung und Forschung
through Grant No.\ 05~HT4GUA/4.
\end{acknowledgments}

\appendix

\allowdisplaybreaks

\boldmath
\section{Coefficients for Subprocess $g+g \to c + X$}
\label{app:gg}
\unboldmath

In this appendix, we list the coefficients $c_i$ needed for the
calculation of the cross section for the inclusive production of
charm in $gg$ collisions. They are obtained by taking the limit $m
\to 0$ of the cross sections calculated in Ref.~\cite{BS}.
These limits are compared with the results from Ref.~\cite{ACGG} in order to
obtain the subtraction terms $\Delta c_i$. To shorten the expressions, we make
use of the abbreviations
\begin{equation}
X = 1 - vw \, , \qquad Y = 1 - v + vw \, , \qquad v_i = i -v \, ,
\end{equation}
and write down only non-zero contributions. The coefficients given below
have to be used together with Eqs.~(\ref{sigma_massless}) and (\ref{subt});
their colour decomposition is defined in Eq.~(\ref{col_decomp}).  We start
with the Abelian coefficients $c^{\rm qed}_{i}$.

\boldmath
\subsection{$c^{\rm qed}_{i}$ Coefficients}
\label{app:ggqed}
\unboldmath

\begin{eqnarray}
c^{\rm qed}_{1} &=&
\frac{-21 - 6 \ln v + 9 \ln^2 v + 
    3 \ln v_1 + 3 \ln^2 v_1 + 4 {\pi }^2
    }{6} \tau(v)
\nonumber\\ & &{} +
  \frac{5 \ln v_1 + \ln^2 v_1 - 
     \ln v + 3 \ln^2 v}{2 v_1}
+
  \frac{5 \ln v + \ln^2 v - 
     \ln v_1 + 3 \ln^2 v_1}{2 v}
+ \Delta c^{\rm qed}_{1} \, ,
\nonumber\\
\Delta c^{\rm qed}_{1} &=& 2(1 -  \ln v - \ln^2 v) \tau(v) \, ,
\nonumber\\
\tilde{c}^{\rm qed}_{1} &=& -\frac{1}{2} ( 3 + 4 \ln v ) \tau(v) \, , 
\nonumber\\
c^{\rm qed}_{2} &=& \frac{1}{2} ( 4 \ln v - 3 ) \tau(v) 
+ \Delta c^{\rm qed}_{2} \, ,
\nonumber\\
\Delta c^{\rm qed}_{2} &=& -2(1 + 2 \ln v) \tau(v) \, ,
\nonumber\\
\tilde{c}^{\rm qed}_{2} &=& -2 \tau(v) \, ,
\nonumber\\
c^{\rm qed}_{3} &=& 2 \tau(v) + \Delta c^{\rm qed}_{3} \, ,
\nonumber\\
\Delta c^{\rm qed}_{3} &=& -4 \tau(v) \, ,
\nonumber\\
c^{\rm qed}_{5} &=&
\frac{-2( 1 + 2v^2 ) }{v} + 
  \frac{3 - 2v + 2v^2}{v w} - 
  \frac{2( -1 + v - v^2 + 2v^3 ) w}{v_1 v} -
  \frac{2( -v + v^2 ) }{X^3}
\nonumber\\
&&{}
- \frac{2v}{X^2} + 
  \frac{3v - 4v^2 + 2v^3}{v_1 X } - 
  \frac{2v}{Y} + \Delta c^{\rm qed}_{5} \, ,
\nonumber\\
\Delta c^{\rm qed}_{5} &=&
\frac{2 v}{v_1} - \frac{2 ( 2 - 2 v + v^2 ) }{v w} +
  \frac{2 v^2 w}{v_1} + \frac{4 v}{Y} \, ,
\nonumber\\
c^{\rm qed}_{6} &=&
\frac{-4 v^2}{v_1} 
+ \frac{2 ( -2 + 5 v - 3 v^2 + 2 v^3 ) }{v_1  v w} 
+ \frac{2 v^2 w}{v_1} \, , 
\nonumber\\
c^{\rm qed}_{7} &=& \frac{-2}{v w} - \frac{4 v}{Y} \, , 
\nonumber\\
c^{\rm qed}_{8} &=&\frac{2 ( 2 - 2 v + v^2 ) }{v w} \, ,
\nonumber\\
c^{\rm qed}_{9} &=&
\frac{-2 v}{v_1} + \frac{2}{v w} -
  \frac{2 v^2 w}{v_1} \, ,
\nonumber\\
c^{\rm qed}_{10} &=&
\frac{2 ( -1 + v - 2 v^2 + 4 v^3 ) }
   {v_1  v} - 
  \frac{-3 + 7 v - 4 v^2 + 4 v^3}
   {v_1  v w} - 
  \frac{2 ( -1 + v - v^2 + 3 v^3 )  w}
   {v_1  v}
\nonumber\\
&&{}
+ 
  \frac{2 ( v - v^2 ) }{X^3} - 
  \frac{2 v}{X^2} + 
  \frac{3 v - 4 v^2 + 2 v^3}{v_1 X } - 
  \frac{2 v}{Y} + \Delta c^{\rm qed}_{10} \, ,
\nonumber\\
\Delta c^{\rm qed}_{10} &=& \Delta c^{\rm qed}_{5} \, ,
\nonumber\\
c^{\rm qed}_{11} &=&
\frac{7 - 8 v + 6 v^2 - 4 v^3}
   {v_1  v} + 
  \frac{1 - 2 v + 4 v^2}{2 v w} +
  \frac{( -8 + 8 v - 3 v^2 + v^3 )  w}
   {v_1  v}
\nonumber\\
&&{}
- \frac{8 v_1  v}{X^3} + 
  \frac{7 v}{X^2} + 
  \frac{v_2 ( -v + 3 v^2 ) }{v_1 X } 
+ \frac{2 v}{Y} + \Delta c^{\rm qed}_{11} \, ,
\nonumber\\
\Delta c^{\rm qed}_{11} &=& \frac{1}{2} \Delta c^{\rm qed}_{10} \, ,
\nonumber\\
\tilde{c}^{\rm qed}_{11} &=&
\frac{2 ( 1 + 2 v^2 ) }{v} + 
  \frac{-3 + 2 v - 2 v^2}{v w} + 
  \frac{2 ( -1 + v - v^2 + 2 v^3 )  w}{v_1  v} 
- \frac{2 v v_1 }{X^3} 
\nonumber\\
&&{}
+ 
  \frac{2 v}{X^2} + 
  \frac{-3 v + 4 v^2 - 2 v^3}{v_1 X} + 
  \frac{2 v}{Y} \, ,
\nonumber\\
c^{\rm qed}_{12} &=&
-4 - \frac{2}{v_1} - \frac{2}{v} \, ,
\nonumber\\
c^{\rm qed}_{13} &=&
-2 + \frac{4}{v_1} + \frac{2}{v} \, ,
\nonumber\\
c^{\rm qed}_{14} &=&
-2 + \frac{2}{v_1} + \frac{4}{v} \, .
\end{eqnarray}

We note that these results for the Abelian part agree with the
coefficients for $\gamma + \gamma \to c + X$ in Ref.~\cite{KS1}, if the
normalization factor $C(s)$ in Ref.~\cite{KS1} is replaced by $C(s)=1$.  In
that work, the zero-mass limit was derived from the FFN cross
sections of Ref.~\cite{KMC} and compared with the ZM-VFN ones of
Ref.~\cite{LG}.

\boldmath
\subsection{$c^{\rm oq}_{i}$ Coefficients}
\label{app:ggoqu}
\unboldmath

\begin{eqnarray}
c^{\rm oq}_{1} &=& 
\biggl[\frac{-1 + 7 \ln v + 16 \ln^2 v + 2 \ln v_1 - 16 \ln v \ln v_1 + 
     4 \ln^2 v_1}{2}
\nonumber\\
& &{}
- \frac{( -30 - 3 \ln v + 120 \ln^2 v - 15 \ln v_1 - 96 \ln v \ln v_1 + 
    36 \ln^2 v_1 + 8 {\pi }^2 )  v}{6}
\nonumber\\
& &{}
+ \frac{( -15 - 9 \ln v + 42 \ln^2 v - 24 \ln v \ln v_1 + 
    12 \ln^2 v_1 + 4 {\pi }^2 )  v^2}{3}\biggr]\tau(v) 
\nonumber\\
& &{}
-
\frac{-1 + 2 \ln v + 4 \ln^2 v + 
     7 \ln v_1}{2 v_1 } + 
  \frac{1 - 7 \ln v - 2 \ln v_1 - 
     4 \ln^2 v_1}{2 v}
+ \Delta c^{\rm oq}_{1} \, ,
\nonumber\\
\Delta c^{\rm oq}_{1} &=& 
4 ( -1 + \ln v + \ln^2 v ) v_1  v \tau(v) \, ,
\nonumber\\
\tilde{c}^{\rm oq}_{1} &=& 
\big[-4 ( \ln v - \ln v_1 )  + 
  ( 3 + 12 \ln v - 8 \ln v_1 )
      v - ( 3 + 12 \ln v - 
     8 \ln v_1 )  v^2\big] \tau(v) \, ,
\nonumber\\
c^{\rm oq}_{2} &=& 
\big[-8 ( -2 \ln v + \ln v_1 )  + 
  ( 3 - 36 \ln v + 16 \ln v_1
     )  v - ( 3 - 28 \ln v + 
     8 \ln v_1 )  v^2\big] \tau(v)
\nonumber\\
& &{}
+ \Delta c^{\rm oq}_{2} \, ,
\nonumber\\
\Delta c^{\rm oq}_{2} &=& 
4 (1 + 2 \ln v) v v_1 \tau(v) \, ,
\nonumber\\
\tilde{c}^{\rm oq}_{2} &=& 
(-8 + 20 v - 20 v^2) \tau(v) \, ,
\nonumber\\
c^{\rm oq}_{3} &=& 
4 ( 4 - 9 v + 9 v^2 ) \tau(v)
+ \Delta c^{\rm oq}_{3} \, ,
\nonumber\\
\Delta c^{\rm oq}_{3} &=& 8 v_1 v \tau(v) \, ,
\nonumber\\
c^{\rm oq}_{5} &=&
\frac{-2 ( -2 + 12 v - 30 v^2 + 81 v^3 - 81 v^4 + 32 v^5 ) }{v_1^2 v} 
+ \frac{8 ( 4 - 13 v + 20 v^2 - 14 v^3 + 4 v^4 ) }{v_1  v w^2} 
\nonumber\\
& &{}
+ 
\frac{2 ( 4 + 19 v - 112 v^2 + 189 v^3 - 128 v^4 + 32 v^5 ) }{v_1^2 v w} 
\nonumber\\
& &{}
+ \frac{4 ( -1 + 6 v - 12 v^2 + 43 v^3 - 52 v^4 + 28 v^5 )  w}{v_1^2 v}
- \frac{48 v^4 w^2}{v_1^2} + \frac{32 v^4 w^3}{v_1^2} 
\nonumber\\
& &{}
- \frac{4 v_1  v}{X^3} + \frac{4 v ( 5 - 4 v ) }{X^2} - 
\frac{4 ( -3 v + 13 v^2 - 13 v^3 + 4 v^4) }{v_1 X } + 
\frac{4 v_1^2 v}{Y^3} 
\nonumber\\
& &{}
- \frac{4 v_1  v ( 6 - 5 v ) }{Y^2} 
- \frac{2 ( -5 v - 4 v^2 + 4 v^3 ) }{Y}
+ \Delta c^{\rm oq}_{5} \, ,
\nonumber\\
\Delta c^{\rm oq}_{5} &=&
-4 v + \frac{8 v v_1^2 }{Y^3} - 
\frac{8 v^2 v_1 }{Y^2} + 
\frac{4 v( 3  - 6 v + 4 v^2 ) }{Y} \, ,
\nonumber\\
c^{\rm oq}_{6} &=&
\frac{2 v ( 13 - 5 v + 8 v^2 ) }{v_1^2} - 
  \frac{8 ( 1 - 2 v + 2 v^2 ) }{v_1  v w^2} + 
  \frac{4 ( -1 - 2 v + 10 v^2 - 13 v^3 + 4 v^4 ) }{v_1^2 v w} 
\nonumber\\
& &{}
-\frac{8 v^2 ( 11 - 11 v + 8 v^2 )  w}{v_1^2} + 
 \frac{32 v^3 ( 1 + v )  w^2}{v_1^2} - 
  \frac{32 v^4 w^3}{v_1^2} \, ,
\nonumber\\
c^{\rm oq}_{7} &=& 
\frac{6 v ( -7 + 6 v ) }{v_1} - 
  \frac{2 ( -1 + 4 v ) }{w} + 
  \frac{8 v^2 w}{v_1} 
\nonumber\\
& &{}
+ \frac{8 v v_1^2 }{Y^3} 
- \frac{8 ( 6 v - 11 v^2 + 5 v^3 ) }{Y^2} + 
  \frac{4 ( 25 v - 18 v^2 + 4 v^3 ) }{Y} \, ,
\nonumber\\
c^{\rm oq}_{8} &=&
\frac{-2 v ( 6 + 17 v - 27 v^2 + 16 v^3 ) }{v_1^2} + 
\frac{8 ( 1 - 2 v + 2 v^2 ) }{v_1  w^2} + 
\frac{8 ( 1 - v + v^3 ) }{v_1^2 w} 
\nonumber\\
& &{}
+ 
  \frac{4 v ( 4 + 15 v - 27 v^2 + 20 v^3 )
        w}{v_1^2} - 
  \frac{48 v^4 w^2}{v_1^2} + 
  \frac{32 v^4 w^3}{v_1^2} 
\nonumber\\
& &{}
- \frac{4 ( 7 v - 8 v^2 + 4 v^3 ) }{X} 
+ \frac{12 v_1  v}{Y} \, ,
\nonumber\\
c^{\rm oq}_{9} &=&
\frac{-2 ( 6 - 25 v + 76 v^2 - 77 v^3 + 32 v^4 ) }{v_1^2} - 
\frac{8 ( 2 - 7 v + 12 v^2 - 10 v^3 + 4 v^4 ) }{( -1 + v )  v w^2} 
\nonumber\\
& &{}
+ 
\frac{2 ( 2 + 9 v - 54 v^2 + 101 v^3 - 78 v^4 + 24 v^5 ) }{v_1^2 v w} 
+ \frac{8 v^2 ( 9 - 11 v + 6 v^2 )  w}{v_1^2} 
\nonumber\\
& &{}
- \frac{16 v^2 ( 1 - v + v^2 )  w^2}{v_1^2} + 
 \frac{4 ( 5 v - 8 v^2 + 4 v^3 ) }{v_1 X } 
- \frac{12 v_1  v}{Y} \, ,
\nonumber\\
c^{\rm oq}_{10} &=&
\frac{2 ( 2 - 6 v + 11 v^2 - 47 v^3 + 48 v^4 - 16 v^5 ) }{v_1^2 v} 
+ \frac{8 ( 3 - 10 v + 16 v^2 - 12 v^3 + 4 v^4 ) }{v_1  v w^2} 
\nonumber\\
& &{}
+ 
\frac{2 ( 4 + 11 v - 76 v^2 + 135 v^3 - 94 v^4 + 24 v^5 ) }{v_1^2 v w} 
\nonumber\\
& &{}
+ \frac{4 ( -1 + 6 v - 12 v^2 + 41 v^3 - 50 v^4 + 24 v^5 )  w}{v_1^2 v}
- \frac{16 v^2 ( -1 + v + 2 v^2 )  w^2}{v_1^2} 
\nonumber\\
& &{}
+ \frac{32 v^4 w^3}{v_1^2} - \frac{4 v v_1 }{X^3} 
- \frac{4 v( -5  + 4 v ) }{X^2} 
- \frac{4 ( 2 v + 5 v^2 - 9 v^3 + 4 v^4 ) }{v_1 X } 
\nonumber\\
& &{}
+ \frac{4 v v_1^2 }{Y^3} 
+ \frac{4 v_1 v( -6  + 5 v ) }{Y^2} 
- \frac{2 v( -11  + 2 v + 4 v^2 ) }{Y}
+ \Delta c^{\rm oq}_{10} \, ,
\nonumber\\
\Delta c^{\rm oq}_{10} &=& \Delta c^{\rm oq}_{5} \, ,
\nonumber\\
c^{\rm oq}_{11} &=&
\frac{-10 + 33 v - 26 v^2 + 57 v^3 - 94 v^4 + 48 v^5}{v_1^2 v} 
+ \frac{8 ( 3 - 4 v + 2 v^2 ) }{w^2} 
\nonumber\\
& &{}
- \frac{2 ( -1 + 14 v - 30 v^2 + 16 v^3 ) }{v w} 
- \frac{2 ( -6 + 21 v - 23 v^2 + 41 v^3 - 61 v^4 + 32 v^5 )  w}{v_1^2 v} 
\nonumber\\
& &{}
+ \frac{16 v^3 ( -1 + 2 v )  w^2}{v_1^2} 
- \frac{16 v^4 w^3}{v_1^2} 
+ \frac{12 v_1  v}{X^3} 
+ \frac{2 v ( -14 + 9 v ) }{X^2} 
+ \frac{13 v - 9 v^2 - 2 v^3}{v_1 X} 
\nonumber\\
& &{}
- \frac{16 v_1^2 v}{Y^3} 
+ \frac{4 v_1  v (13 - 9 v)}{Y^2} 
+ \frac{-65 v + 56 v^2 - 10 v^3}{Y}
+ \Delta c^{\rm oq}_{11} \, ,
\nonumber\\
\Delta c^{\rm oq}_{11} &=& \frac{1}{2} \Delta c^{\rm oq}_{10} \, ,
\nonumber\\
\tilde{c}^{\rm oq}_{11} &=&
\frac{2 ( -2 + 12 v - 26 v^2 + 69 v^3 - 73 v^4 + 32 v^5 ) }{v_1^2 v} 
- \frac{8 ( 2 - 7 v + 12 v^2 - 10 v^3 + 4 v^4 ) }{v_1  v w^2} 
\nonumber\\
& &{}
- \frac{2 ( 17 - 68 v + 113 v^2 - 82 v^3 + 24 v^4 ) }{v_1^2 w} 
+ \frac{4 ( 1 - 6 v + 12 v^2 - 43 v^3 + 52 v^4 - 28 v^5 )  w}{v_1^2 v} 
\nonumber\\
& &{}
+ \frac{48 v^4 w^2}{v_1^2} 
- \frac{32 v^4 w^3}{v_1^2} 
+ \frac{4 v_1  v}{X^3} 
+ \frac{4 v ( -5 + 4 v ) }{X^2} 
+ \frac{4 v ( 9 - 10 v + 3 v^2 ) }{v_1 X} 
\nonumber\\
& &{}
- \frac{4 v_1^2 v}{Y^3} 
+ \frac{4 v_1  v (6 - 5 v)}{Y^2} 
- \frac{2 ( 25 v - 18 v^2 + 4 v^3 ) }{Y} \, ,
\nonumber\\
c^{\rm oq}_{12} &=& 
16 ( 1 - v + v^2 ) \, ,
\nonumber\\
c^{\rm oq}_{13} &=&
-16 
+ 24 v - 16 v^2
- \frac{4}{v_1} 
+ \frac{4}{v} \, ,
\nonumber\\
c^{\rm oq}_{14} &=&
-32 
+ 40 v - 32 v^2
+ \frac{4}{v_1} 
+ \frac{4}{v} \, .
\end{eqnarray}

\boldmath
\subsection{$c^{\rm kq}_{i}$ Coefficients}
\label{app:ggkqu}
\unboldmath

\begin{eqnarray}
c^{\rm kq}_{1} &=&
\biggl[\frac{-1 - 7 \ln v - 17 \ln^2 v - 6 \ln v_1 + 16 \ln v \ln v_1}{2} 
\nonumber\\
& &{}
+ \frac{( -54 - 15 \ln v + 15 \ln^2 v - 3 \ln v_1 
- 3 \ln^2 v_1 + 8 {\pi }^2 )  v}{6}
\nonumber\\
& &{}
- \frac{( -27 - 9 \ln v + 6 \ln^2 v + 4 {\pi }^2 )  v^2}{3}\biggr] \tau(v)
\nonumber\\
& &{}
+\frac{1 + 7 \ln v_1 + \ln^2 v_1 + 6 \ln v + 8 \ln^2 v }{2 v_1 } 
+ \frac{1 + 7 \ln v + \ln^2 v + 6 \ln v_1 + 8 \ln^2 v_1}{2 v}
+\Delta c^{\rm kq}_{1} \, ,
\nonumber\\
\Delta c^{\rm kq}_{1} &=& 
4 ( 1 - \ln v - \ln^2 v ) v v_1 \tau(v) \, ,
\nonumber\\
\tilde{c}^{\rm kq}_{1} &=& 
\big[4 ( \ln v - \ln v_1 )  - 
  ( 3 + 4 \ln v )  v v_1 \big] \tau(v) \, ,
\nonumber\\
c^{\rm kq}_{2} &=& 
\big[-16 \ln v + ( -3 + 4 \ln v ) v v_1 \big] \tau(v)
+ \Delta c^{\rm kq}_{2} \, ,
\nonumber\\
\Delta c^{\rm kq}_{2} &=&
-4 ( 1 + 2 \ln v )  v v_1 \tau(v) \, ,
\nonumber\\
\tilde{c}^{\rm kq}_{2} &=& 
(8 - 4 v + 4 v^2) \tau(v) \, ,
\nonumber\\
c^{\rm kq}_{3} &=& 
(-16 + 4 v - 4 v^2)\tau(v)
+\Delta c^{\rm kq}_{3} \, ,
\nonumber\\
\Delta c^{\rm kq}_{3} &=& 
-8 v v_1 \tau(v) \, ,
\nonumber\\
c^{\rm kq}_{5} &=&
\frac{-2 ( 2 - 8 v + 6 v^2 - 9 v^3 + v^4 ) }{v_1^2 v} 
- \frac{8 ( 1 - 2 v + 2 v^2 ) }{v_1  v w^2} 
- \frac{2 ( 4 - 5 v + 5 v^3 ) }{v_1^2 v w} 
\nonumber\\
& &{}
- \frac{4 ( -1 + 4 v - 4 v^2 + 5 v^3 )  w}{v_1^2 v} 
+ \frac{4 v_1  v}{X^3} 
+ \frac{4 v ( -3 + 2 v ) }{X^2} 
- \frac{4 v ( -5 + 2 v ) }{v_1 X} 
\nonumber\\
& &{}
+ \frac{4 v_1^2 v}{Y^3} 
- \frac{4 v_1  v^2}{Y^2} 
+ \frac{2 v ( 3 - 6 v + 4 v^2 ) }{Y}
+ \Delta c^{\rm kq}_{5} \, ,
\nonumber\\
\Delta c^{\rm kq}_{5} &=&
4 v - \frac{8 v v_1^2 }{Y^3} 
+ \frac{8 v^2 v_1 }{Y^2} 
- \frac{4 ( 3 v - 6 v^2 + 4 v^3 ) }{Y} \, ,
\nonumber\\
c^{\rm kq}_{6} &=&
\frac{2 v ( -5 - 5 v + 2 v^2 ) }{v_1^2} 
+ \frac{8 ( 1 - 2 v + 2 v^2 ) }{v_1  v w^2} 
- \frac{2 ( 2 - 13 v + 16 v^2 - 11 v^3 + 2 v^4 ) }{v_1^2 v w} 
\nonumber\\
& &{}
-\frac{2 ( -9 + v )  v^2 w}{v_1^2} \, ,
\nonumber\\
c^{\rm kq}_{7} &=& 
-6 v - \frac{2 ( 4 - v + v^2 ) }{v w} 
+ \frac{2 v^2 w}{v_1} + \frac{8 v_1^2 v}{Y^3} 
- \frac{8 v_1  v^2}{Y^2} + 
\frac{4 v ( 3 - 6 v + 4 v^2 ) }{Y} \, ,
\nonumber\\
c^{\rm kq}_{8} &=&
\frac{2 v ( 1 + 7 v ) }{v_1^2} 
- \frac{8 ( 1 - 2 v + 2 v^2 ) }{v_1  v w^2} 
+ \frac{2 ( 4 - 14 v + 17 v^2 - 12 v^3 + v^4) }{v_1^2 v w} 
\nonumber\\
& &{}
+ \frac{4 v ( 1 - 5 v )  w}{v_1^2} \, ,
\nonumber\\
c^{\rm kq}_{9} &=&
\frac{2 v^2 ( 7 + v ) }{v_1^2} 
- \frac{8 ( 1 - 2 v + 2 v^2 ) }{v_1  v w^2} 
- \frac{2 ( 2 - 3 v + 2 v^2 + v^3 + 2 v^4 )}{v_1^2 v w} 
\nonumber\\
& &{}
+ \frac{2 ( -9 + v )  v^2 w}{v_1^2} 
+ \frac{4 v ( 2 - 2 v + v^2 ) }{v_1 X} \, ,
\nonumber\\
c^{\rm kq}_{10} &=&
\frac{-2 ( 2 - 8 v + 4 v^2 - 9 v^3 + 3 v^4) }{v_1^2 v} 
- \frac{8 ( 1 - 2 v + 2 v^2 ) }{v_1  v w^2} 
\nonumber\\
& &{}
+ \frac{2 ( -4 + 2 v + 3 v^2 - 8 v^3 + 3 v^4) }{v_1^2 v w} 
+ \frac{2 ( 2 - 8 v + 8 v^2 - 11 v^3 + v^4 )w}{v_1^2 v} 
\nonumber\\
& &{}
+ \frac{4 v_1  v}{X^3} 
+ \frac{4 v ( -3 + 2 v ) }{X^2} 
- \frac{4 v ( -3 + v^2 ) }{v_1 X} 
\nonumber\\
& &{}
+ \frac{4 v_1^2 v}{Y^3} 
- \frac{4 v_1  v^2}{Y^2} 
+ \frac{2 v ( 3 - 6 v + 4 v^2 ) }{Y}
+ \Delta c^{\rm kq}_{10} \, ,
\nonumber\\
\Delta c^{\rm kq}_{10} &=& \Delta c^{\rm kq}_{5} \, ,
\nonumber\\
c^{\rm kq}_{11} &=&
- \frac{-10 + 31 v - 22 v^2 + 7 v^3 + 2 v^4}{v_1^2 v}   
- \frac{8}{w^2} 
- \frac{2 ( 1 - 6 v + 6 v^2 ) }{v w} 
\nonumber\\
& &{}
- \frac{2 ( 2 - 5 v + v^2 ) ( 3 - 2 v + v^2 )  w}{v_1^2 v} 
- \frac{12 v_1  v}{X^3} 
- \frac{2 v ( -12 + 7 v ) }{X^2} 
+ \frac{v ( -3 - 9 v + 10 v^2 ) }{v_1 X} 
\nonumber\\
& &{}
- \frac{8 v_1^2 v}{Y^3} 
+ \frac{4 v_1  v ( -1 + 3 v ) }{Y^2} 
- \frac{v ( -19 + 6 v^2 ) }{Y}
+ \Delta c^{\rm kq}_{11} \, ,
\nonumber\\
\Delta c^{\rm kq}_{11} &=& \frac{1}{2} \Delta c^{\rm kq}_{10} \, ,
\nonumber\\
\tilde{c}^{\rm kq}_{11} &=&
\frac{2 ( 2 - 8 v + 10 v^2 - 13 v^3 + v^4 )}{v_1^2 v} 
+ \frac{8 ( 1 - 2 v + 2 v^2 ) }{v_1  v w^2} 
- \frac{2 ( -5 + 12 v - 13 v^2 + 2 v^3 ) }{v_1^2 w} 
\nonumber\\
& &{}
+ \frac{4 ( -1 + 4 v - 4 v^2 + 5 v^3 )  w}{v_1^2 v} 
- \frac{4 v_1  v}{X^3} 
- \frac{4 v ( -3 + 2 v ) }{X^2} 
+ \frac{4 v ( -3 + v^2 ) }{v_1 X} 
\nonumber\\
& &{}
- \frac{4 v_1^2 v}{Y^3} 
+ \frac{4 v_1  v^2}{Y^2} 
- \frac{2 v ( 3 - 6 v + 4 v^2 ) }{Y} \, ,
\nonumber\\
c^{\rm kq}_{12} &=&
6 - 16 v + 16 v^2
- \frac{8}{v_1} - \frac{8}{v} \, ,
\nonumber\\
c^{\rm kq}_{13} &=&
10 - 2 v
+ \frac{4}{v_1} - \frac{4}{v} \, ,
\nonumber\\
c^{\rm kq}_{14} &=&
8 + 2 v
- \frac{4}{v_1} + \frac{4}{v} \, .
\end{eqnarray}

\subsection{Coefficient for the Quark-Loop Contribution}
\label{app:ggql}

\begin{eqnarray}
c_1^{\rm ql} &=& \Delta c_1^{\rm ql} \, ,
\nonumber\\
\Delta c_1^{\rm ql} &=& -\frac{4}{9} v(1-v) \, .
\end{eqnarray}

\boldmath
\section{Coefficients for Subprocess $q+\bar{q} \to c + \bar{c}$}
\label{app:qq}
\unboldmath

In the following, we list the coefficients $c_i$ needed for the
calculation of the cross section for the inclusive production of
charm in $q \bar{q}$ collisions.  As before, the coefficients $c_i$
are obtained from the cross sections of Ref.~\cite{BS} by
taking the limit $m \to 0$, whereas the subtraction terms $\Delta c_i$
are deduced by a comparison with the results of Ref.~\cite{ACGG}. The
coefficients given in this appendix determine the cross section
according to Eqs.~(\ref{sigma_massless}) and (\ref{subt}) and using the color
decomposition of Eq.~(\ref{col}).  We start with the Abelian coefficients
$c^{\rm cf}_{i}$.

\boldmath
\subsection{$c^{\rm cf}_{i}$ Coefficients}
\label{app:qqcf}
\unboldmath

\begin{eqnarray}
c^{\rm cf}_{1} &=&
\frac{1}{3}\big[24 v \ln v_1  
+ 6 ( 3 - 6 v + 2 v^2 )  \ln^2 v_1 
+ 5 ( -15 + 2 {\pi }^2 )  ( 1 - 2 v + 2 v^2 )  
\nonumber\\
& &{}
- 24 ( 3 - 6 v + 7 v^2 )  \ln^2 v 
+ 3 
( -11 + 14 v - 6 v^2 
+ 28 ( 1 - 2 v + 2 v^2 ) \ln v_1 ) 
\ln v 
\big]
\nonumber\\
& &{}
+ \Delta c^{\rm cf}_{1} \, ,
\nonumber\\
\Delta c^{\rm cf}_{1} &=& 
-4 ( -1 + \ln v + \ln^2 v ) \tau_q(v) \, ,
\nonumber\\
\tilde{c}^{\rm cf}_{1} &=& 
(-9 - 8 \ln v + 4 \ln v_1) \tau_q(v) \, ,
\nonumber\\
c^{\rm cf}_{2} &=& 
(-3 - 20 \ln v + 24 \ln v_1) \tau_q(v)
+ \Delta c^{\rm cf}_{2} \, ,
\nonumber\\
\Delta c^{\rm cf}_{2} &=& 
-4 ( 1 + 2 \ln v )  \tau_q(v) \, ,
\nonumber\\
\tilde{c}^{\rm cf}_{2} &=& -12 \tau_q(v) \, ,
\nonumber\\
c^{\rm cf}_{3} &=& 20 \tau_q(v)
+ \Delta c^{\rm cf}_{3} \, ,
\nonumber\\
\Delta c^{\rm cf}_{3} &=& -8 \tau_q(v) \, ,
\nonumber\\
c^{\rm cf}_{5} &=&
\frac{2 ( -1 + 34 v - 39 v^2 + 8 v^3 ) }{v_1} 
+ \frac{4 ( 1 - 2 v + 2 v^2 ) }{w} 
- \frac{8 v^2 w}{v_1} 
+ \frac{8 v^3 w^2}{v_1} 
- \frac{2 v}{X} 
\nonumber\\
& &{}
+ \frac{4 v_1^2 v}{Y^3} 
- \frac{4 v_1  v^2}{Y^2} 
- \frac{2 ( 55 v - 74 v^2 + 28 v^3 ) }{Y}
+ \Delta c^{\rm cf}_{5} \, ,
\nonumber\\
\Delta c^{\rm cf}_{5} &=&
4 v - \frac{8 v v_1^2 }{Y^3} 
+ \frac{8 v^2 v_1}{Y^2} 
- \frac{4 ( 3 v - 6 v^2 + 4 v^3 ) }{Y} \, ,
\nonumber\\
c^{\rm cf}_{6} &=&
\frac{-4 v ( 7 - 10 v + 4 v^2 ) }{v_1} + 
  \frac{8 (2 - v )  v^2 w}{v_1} 
- \frac{8 v^3 w^2}{v_1} \, ,
\nonumber\\
c^{\rm cf}_{7} &=& 
-20 v + \frac{8 v v_1^2 }{Y^3} 
- \frac{8 v^2 v_1 }{Y^2} 
+ \frac{4 ( v - 6 v^2 + 4 v^3 ) }{Y} \, ,
\nonumber\\
c^{\rm cf}_{8} &=&
\frac{4 v ( -3 + 2 v + 2 v^2 ) }{v_1} 
- \frac{8 v^2 w}{v_1} 
+ \frac{8 v^3 w^2}{v_1} 
+ \frac{16 ( 5 v - 8 v^2 + 4 v^3 ) }{Y} \, ,
\nonumber\\
c^{\rm cf}_{9} &=&
40 v + \frac{4 ( 1 - 2 v + 2 v^2 ) }{w} 
+ 8 v^2 w 
- \frac{16 ( 5 v - 8 v^2 + 4 v^3 ) }{Y} \, ,
\nonumber\\
c^{\rm cf}_{10} &=&
\frac{2 ( -1 + 22 v - 31 v^2 + 12 v^3 ) }{v_1} 
+ \frac{4 ( 1 - 2 v + 2 v^2 ) }{w} 
- \frac{8 ( 2 - v )  v^2 w}{v_1} 
+ \frac{8 v^3 w^2}{v_1} 
- \frac{2 v}{X} 
\nonumber\\
& &{}
+ \frac{4 v v_1^2}{Y^3} 
- \frac{4 v^2 v_1 }{Y^2} + 
  \frac{2 ( -15 v + 10 v^2 + 4 v^3 ) }{Y}
+ \Delta c^{\rm cf}_{10} \, ,
\nonumber\\
\Delta c^{\rm cf}_{10} &=& \Delta c^{\rm cf}_{5} \, ,
\nonumber\\
c^{\rm cf}_{11} &=&
\frac{-10 + 9 v + 3 v^2}{v_1}   
+ \frac{2}{w} 
+ \frac{6 v}{X} 
- \frac{8 v v_1^2 }{Y^3} 
- \frac{12 ( 3 v - 4 v^2 + v^3 ) }{Y^2} 
- \frac{2 ( -11 v + 3 v^3 ) }{Y}
\nonumber\\
& &{}
+ \Delta c^{\rm cf}_{11} \, ,
\nonumber\\
\Delta c^{\rm cf}_{11} &=& \frac{1}{2} \Delta c^{\rm cf}_{10} \, ,
\nonumber\\
\tilde{c}^{\rm cf}_{11} &=&
\frac{2 ( 1 - 10 v + 15 v^2 - 8 v^3 ) }{v_1} 
- \frac{4 ( 1 - 2 v + 2 v^2 ) }{w} 
+ \frac{8 v^2 w}{v_1} 
- \frac{8 v^3 w^2}{v_1} 
+ \frac{2 v}{X} 
\nonumber\\
& &{}
- \frac{4 v_1^2 v}{Y^3} 
+ \frac{4 v_1  v^2}{Y^2} 
- \frac{2 ( v - 6 v^2 + 4 v^3 ) }{Y} \, ,
\nonumber\\
c^{\rm cf}_{12} &=&
8 ( -1 + 2 v + 2 v^2 ) \, ,
\nonumber\\
c^{\rm cf}_{13} &=&
-8 ( 3 - 6 v + 8 v^2 ) \, ,
\nonumber\\
c^{\rm cf}_{14} &=&
16 v_1^2 \, .
\end{eqnarray}

\boldmath
\subsection{$c^{\rm ca}_{i}$ Coefficients}
\label{app:qqca}
\unboldmath

\begin{eqnarray}
c^{\rm ca}_{1} &=&
\frac{1}{9}\big[-18 v \ln v_1 + 
    9 ( -1 + 2 v )  \ln^2 v_1 
- 2 ( -85 + 9 {\pi }^2 )  ( 1 - 2 v + 2 v^2 )  
\nonumber\\
& &{ }
+ 18 ( 7 - 14 v + 16 v^2 )  \ln^2 v 
- 36 ( -1 + v + ( 3 - 6 v + 6 v^2 )  \ln v_1 ) \ln v 
\big] \, ,
\nonumber\\
\tilde{c}^{\rm ca}_{1} &=& \frac{22}{3} \tau_q(v) \, ,
\nonumber\\
c^{\rm ca}_{2} &=& 
16 \tau_q(v) \ln v \, ,
\nonumber\\
\tilde{c}^{\rm ca}_{2} &=& 0 \, ,
\nonumber\\
c^{\rm ca}_{3} &=& 0 \, ,
\nonumber\\
c^{\rm ca}_{5} &=&
-4 v ( 4 + v )  - 4 v^2 w 
- \frac{4 v v_1^2}{Y^3} 
+ \frac{4 ( 3 v - 5 v^2 + 2 v^3 ) }{Y^2} 
+ \frac{2 ( 19 v - 28 v^2 + 12 v^3 ) }{Y} \, ,
\nonumber\\
c^{\rm ca}_{6} &=& 4 v \, ,
\nonumber\\
c^{\rm ca}_{7} &=& 
18 v 
- \frac{8 v v_1^2}{Y^3} 
+ \frac{8 ( 3 v - 5 v^2 + 2 v^3 ) }{Y^2} 
- \frac{4 ( 9 v - 12 v^2 + 4 v^3 ) }{Y} \, ,
\nonumber\\
c^{\rm ca}_{8} &=&
-2 v ( -7 + 2 v )  - 4 v^2 w - 
  \frac{2 ( 17 v - 26 v^2 + 12 v^3 ) }
   {Y} \, ,
\nonumber\\
c^{\rm ca}_{9} &=&
-4 v ( 4 + v )  - 4 v^2 w + 
  \frac{2 ( 17 v - 26 v^2 + 12 v^3 ) }
   {Y} \, ,
\nonumber\\
c^{\rm ca}_{10} &=&
-6 v 
- \frac{4 v v_1^2 }{Y^3} 
+ \frac{4 ( 3 v - 5 v^2 + 2 v^3 ) }{Y^2} 
+ \frac{4 v v_1 }{Y} \, ,
\nonumber\\
c^{\rm ca}_{11} &=&
2 ( 2 - 5 v + 2 v^2 )  + 4 v^2 w 
- \frac{2 v}{X} 
+ \frac{12 v v_1^2 }{Y^3} 
- \frac{12 v v_1^2 }{Y^2} 
+ \frac{4 ( 4 v - 7 v^2 + 2 v^3 ) }{Y} \, ,
\nonumber\\
\tilde{c}^{\rm ca}_{11} &=&
4 ( -2 + v )  v + 4 v^2 w 
+ \frac{4 v v_1^2 }{Y^3} 
- \frac{4 ( 3 v - 5 v^2 + 2 v^3 ) }{Y^2} 
+ \frac{2 ( 9 v - 12 v^2 + 4 v^3 ) }{Y} \, ,
\nonumber\\
c^{\rm ca}_{12} &=&
-2 + 4 v - 16 v^2 \, ,
\nonumber\\
c^{\rm ca}_{13} &=&
4 ( 3 - 6 v + 8 v^2 ) \, ,
\nonumber\\
c^{\rm ca}_{14} &=&
-2 + 4 v \, .
\end{eqnarray}

\subsection{Coefficient for the Quark-Loop Contribution}
\label{app:qqql}

\begin{eqnarray}
c_1^{\rm ql} &=& -\frac{20}{9} n_f \tau_q(v) \, ,
\nonumber\\
\tilde{c}_1^{\rm ql} &=& -\frac{4}{3} n_f \tau_q(v) \, .
\end{eqnarray}

\boldmath
\section{Coefficients for Subprocess $g+q \to c + \bar{c}+q$}
\label{app:gq}
\unboldmath

This appendix contains the coefficients needed for the calculation of
the cross section for $g+q \to c+\bar{c}+q$ with an observed charm quark
in the final state according to Eq.~(\ref{sigma_massless}).  We note that
there are no subtraction terms for the $gq$ channel. In the following,
again only non-zero coefficients are written down. According to the
color decomposition defined in Eq.~(\ref{eq:gq-col}), we present the Abelian
and non-Abelian coefficients $c^{\rm cf}_{i}$ and $c^{\rm ca}_{i}$ separately.

\boldmath
\subsection{$c^{\rm cf}_{i}$ Coefficients}
\label{app:gqcf}
\unboldmath

\begin{eqnarray}
c^{\rm cf}_{5} &=&
\frac{-2 ( 1 - 2 v + 6 v^2 - 5 v^3 + 2 v^4 )}{v_1^2} 
+ \frac{2 ( 1 - 2 v + 4 v^2 - 3 v^3 + v^4 ) }{v_1^2 w} 
\nonumber\\
& &{}
+ \frac{2 ( 1 - 2 v + 6 v^2 - 5 v^3 + 2 v^4)  w}{v_1^2} 
+ \frac{v}{2 X^2} 
+ \frac{-3 v + v^2}{2 v_1 X} 
- \frac{v_1  v}{2 Y^2} 
+ \frac{3 v - 2 v^2}{2 Y } \, ,
\nonumber\\
c^{\rm cf}_{6} &=&
\frac{2 v ( 1 + v^2 ) }{v_1^2} 
+ \frac{-1 - v^2}{v_1^2 w} 
- \frac{4 v^3 w}{v_1^2} \, ,
\nonumber\\
c^{\rm cf}_{7} &=& 
\frac{-2 v}{v_1} 
+ \frac{4 v^2 w}{v_1} 
- \frac{v v_1}{Y^2} 
+ \frac{3 v - 2 v^2}{Y} \, ,
\nonumber\\
c^{\rm cf}_{8} &=&
\frac{-2 v ( 3 - 3 v + 2 v^2 ) }{v_1^2} 
+ \frac{1 + v^2}{v_1^2 w} 
+ \frac{4 v ( 2 - 3 v + 2 v^2 )  w}{v_1^2} \, ,
\nonumber\\
c^{\rm cf}_{9} &=&
\frac{-4 ( 1 - 2 v + 5 v^2 - 4 v^3 + v^4 ) }{v_1^2} 
+ \frac{2 ( 1 - 2 v + 4 v^2 - 3 v^3 + v^4 ) }{v_1^2 w} 
\nonumber\\
& &{}
+ \frac{4 v^2 ( 5 - 5 v + v^2 )  w}{v_1^2} 
- \frac{16 v^2 w^2}{v_1} \, ,
\nonumber\\
c^{\rm cf}_{10} &=&
\frac{-2 ( -1 + 2 v - v^2 + 2 v^4 ) }{v_1^2} 
+ \frac{2 ( 1 - 2 v + 4 v^2 - 3 v^3 + v^4 ) }{v_1^2 w} 
\nonumber\\
& &{}
+ \frac{2 ( 1 - 2 v - 4 v^2 + 5 v^3 + 2 v^4)  w}{v_1^2} 
+ \frac{16 v^2 w^2}{v_1} 
+ \frac{v}{2 X^2} 
+ \frac{-3 v + v^2}{2 v_1 X} 
\nonumber\\
& &{}
- \frac{v_1  v}{2 Y^2} 
+ \frac{3 v - 2 v^2}{2 Y } \, ,
\nonumber\\
c^{\rm cf}_{11} &=&
\frac{2 v^2 ( 4 - 5 v + 2 v^2 ) }{v_1^2} 
+ \frac{1 + 2 v - 2 v^2}{2 w} 
\nonumber\\
& &{}
- \frac{( -2 + 4 v + 7 v^2 - 11 v^3 + 4 v^4 )w}{v_1^2} 
- \frac{v}{2 X^2} 
+ \frac{5 v - 4 v^2}{2 v_1 X } 
+ \frac{v_1  v}{Y^2} 
+ \frac{-8 v + 3 v^2}{2 Y } \, ,
\nonumber\\
\tilde{c}^{\rm cf}_{11} &=&
\frac{2 ( 1 - 2 v + 6 v^2 - 5 v^3 + 2 v^4 )}{v_1^2} 
- \frac{2 ( 1 - 2 v + 4 v^2 - 3 v^3 + v^4 ) }{v_1^2 w} 
\nonumber\\
& &{}
- \frac{2 ( 1 - 2 v + 6 v^2 - 5 v^3 + 2 v^4)  w}{v_1^2} 
- \frac{v}{2 X^2} 
+ \frac{3 v - v^2}{2 v_1 X } 
+ \frac{v_1 v}{2 Y^2} 
+ \frac{-3 v + 2 v^2}{2 Y } \, .
\end{eqnarray}

We have checked that the following relations between these results
for the $C_F$ part and the coefficients for the subprocess
$\gamma+q \to c+\bar{c}+q$ in Ref.~\cite{KS2} are satisfied, if
the normalization factors $C(s)$, $C_q(s)$, $C_{cq}(s)$ and
$C_F$ are replaced by unity:
$c^{\rm cf}_{i} = c_i^{Q_1}  + c_i^{Q_2} - 2 c_i^{Q_3}$.
Furthermore, our results in Appendix \ref{app:gqbarcf}
fulfill the relations
$c^{{\rm cf},\bar{c} - c}_{i} = 4 c_i^{Q_3}$ and
$c^{\rm cf}_{i} + c^{{\rm cf},\bar{c} - c}_{i}/2 = c_i^{Q_1}  + c_i^{Q_2}$.

The decomposition of the form
$e_c^2 Q_1 + e_q^2 Q_2 - e_c e_q Q_3$
for $\gamma q \to Q$ turns into one of the form
$Q_1 + Q_2 - 2 Q_3$ for $g q \to Q$.
The interference part $Q_3$ is anti-symmetric and changes sign
in the analogous subprocesses with an anti-quark $\bar{q}$, {\it i.e.}\
we have
$e_c^2 Q_1 + e_q^2 Q_2 + e_c e_q Q_3$
for $\gamma \bar{q} \to Q$ and
$Q_1 + Q_2 + 2 Q_3$ for $g \bar{q} \to Q$.

\boldmath
\subsection{$c^{\rm ca}_{i}$ Coefficients}
\label{app:gqca}
\unboldmath

\begin{eqnarray}
c^{\rm ca}_{5} &=&
- \frac{v^2 ( 2 - 2 v + v^2 ) }{v_1^2}   
+ \frac{4 v^2 ( 1 - v + v^2 )  w}{v_1^2} 
- \frac{6 v^4 w^2}{v_1^2} 
+ \frac{4 v^4 w^3}{v_1^2} 
- \frac{v}{2 X} 
+ \frac{v}{2 Y } \, ,
\nonumber\\
c^{\rm ca}_{6} &=&
\frac{v ( 1 + v ) }{v_1^2} 
- \frac{v^2 ( 5 - v + 2 v^2 )  w}{v_1^2} 
+ \frac{4 v^3 ( 1 + v )  w^2}{v_1^2} 
- \frac{4 v^4 w^3}{v_1^2} \, ,
\nonumber\\
c^{\rm ca}_{7} &=& 
-v - \frac{v^2 w}{v_1} + \frac{v}{Y} \, ,
\nonumber\\
c^{\rm ca}_{8} &=&
- \frac{v ( -1 + 4 v - 3 v^2 + v^3 ) }{v_1^2}   
+ \frac{2 v ( -1 + 4 v - 3 v^2 + 2 v^3 )  w}{v_1^2} 
- \frac{6 v^4 w^2}{v_1^2} 
+ \frac{4 v^4 w^3}{v_1^2} \, ,
\nonumber\\
c^{\rm ca}_{9} &=&
- \frac{-1 + 2 v - v^2 + v^4}{v_1^2}   
+ \frac{v^2 ( -1 + v + 2 v^2 )  w}{v_1^2} 
- \frac{2 v^2 ( -2 + 2 v + v^2 )  w^2}{v_1^2} \, ,
\nonumber\\
c^{\rm ca}_{10} &=&
-\frac{1 - v + 5 v^2}{v_1} 
+ \frac{v^2 ( 11 - 11 v + 2 v^2 )  w}{v_1^2} 
- \frac{4 v^2 ( 1 - v + v^2 )  w^2}{v_1^2} 
+ \frac{4 v^4 w^3}{v_1^2} 
\nonumber\\
& &{}
- \frac{v}{2 X} 
+ \frac{v}{2 Y } \, ,
\nonumber\\
c^{\rm ca}_{11} &=&
1 + v^2 + \frac{2 ( -1 + v + v^3 )  w}{v_1} 
+ \frac{2 v^3 ( -1 + 2 v )  w^2}{v_1^2} 
- \frac{2 v^4 w^3}{v_1^2} 
- \frac{v}{X} 
+ \frac{v}{2 Y } \, ,
\nonumber\\
\tilde{c}^{\rm ca}_{11} &=&
\frac{v^2 ( 2 - 2 v + v^2 ) }{v_1^2} 
- \frac{4 v^2 ( 1 - v + v^2 )  w}{v_1^2} 
+ \frac{6 v^4 w^2}{v_1^2} 
- \frac{4 v^4 w^3}{v_1^2} 
+ \frac{v}{2 X } 
- \frac{v}{2 Y } \, .
\end{eqnarray}

\boldmath
\section{Coefficients for Subprocess $g+\bar{q} \to c + \bar{c}+\bar{q}$}
\label{app:gqbar}
\unboldmath

The cross sections for the subprocesses $g+q \to c+\bar{c}+q$ and $g+\bar{q}
\to c+\bar{c}+\bar{q}$, where the quark in the initial state is replaced by
an anti-quark, are related, but not identical. The differences 
between the corresponding coefficients, $c_i^{\bar{c} - c}$, will be presented
below. They have to be combined with the coefficients for
$g+q \to c+\bar{c}+q$ given in Appendix~\ref{app:gq} to give those for
$g+\bar{q}\to c+\bar{c}+\bar{q}$, according to
$c_i(g+\bar{q} \to c+\bar{c}+\bar{q}) =
c_i(g+q \to c+\bar{c}+q) + c_i^{\bar{c} - c} $, to be inserted in
Eqs.~(\ref{sigma_massless}) and (\ref{eq:gq-col}).
Again, there are no subtraction terms, and we
present only the non-zero coefficients. Note, in particular, that for both
colour factors $c_{1-5}^{\bar{c} - c}$ and $\tilde{c}_{12-14}^{\bar{c} -
  c}$ are zero.

\boldmath
\subsection{$c^{\rm cf}_{i}$ Coefficients}
\label{app:gqbarcf}
\unboldmath

\begin{eqnarray}
c^{{\rm cf},\bar{c} - c}_{6} &=& \frac{4 v^2}{v_1} - \frac{8 v^2 w}{v_1} \, ,
\nonumber\\
c^{{\rm cf},\bar{c} - c}_{7} &=& c^{{\rm cf},\bar{c} - c}_{6} \, ,
\nonumber\\
c^{{\rm cf},\bar{c} - c}_{8} &=& 8 v - 16 v w \, ,
\nonumber\\
c^{{\rm cf},\bar{c} - c}_{9} &=& \frac{4 ( 2 - 2 v + 3 v^2 ) }{v_1} 
- \frac{24 v^2 w}{v_1} 
+ \frac{32 v^2 w^2}{v_1} \, ,
\nonumber\\ 
c^{{\rm cf},\bar{c} - c}_{10} &=&\frac{-4 ( 2 - 2 v + 5 v^2 ) }{v_1} 
+ \frac{40 v^2 w}{v_1} - \frac{32 v^2 w^2}{v_1} \, ,
\nonumber\\
c^{{\rm cf},\bar{c} - c}_{11} &=&8 - 16 w - \frac{4 v}{X} 
+ \frac{4 v}{Y} \, .
\end{eqnarray}

\boldmath
\subsection{$c^{\rm ca}_{i}$ Coefficients}
\label{app:gqbarca}
\unboldmath

\begin{eqnarray}
c^{{\rm ca},\bar{c} - c}_{6} &=& \frac{-3 v^2}{2 v_1 } 
+ \frac{3 v^2 w}{v_1} \, ,
\nonumber\\
c^{{\rm ca},\bar{c} - c}_{7} &=& c^{{\rm ca},\bar{c} - c}_{6} \, ,
\nonumber\\
c^{{\rm ca},\bar{c} - c}_{8} &=& -3 v + 6 v w \, ,
\nonumber\\
c^{{\rm ca},\bar{c} - c}_{9} &=& \frac{-3 ( 2 - 2 v + 3 v^2 ) }{2 v_1} 
+ \frac{9 v^2 w}{v_1} 
- \frac{12 v^2 w^2}{v_1} \, ,
\nonumber\\
c^{{\rm ca},\bar{c} - c}_{10} &=& \frac{3 ( 2 - 2 v + 5 v^2 ) }{2 v_1} 
- \frac{15 v^2 w}{v_1} + \frac{12 v^2 w^2}{v_1} \, ,
\nonumber\\ 
c^{{\rm ca},\bar{c} - c}_{11} &=& -3 + 6 w + \frac{3 v}{2 X} 
- \frac{3 v}{2 Y} \, .
\end{eqnarray}

\end{document}